\documentclass[a4paper, amsfonts, amssymb, amsmath, reprint, showkeys, nofootinbib, twoside, floatfix, pre,superscriptaddress]{revtex4-2}

\usepackage[utf8]{inputenc}
\usepackage[left=23mm,right=13mm,top=35mm,columnsep=15pt]{geometry} 
\usepackage{amsmath}
\usepackage{amssymb}
\usepackage{physics}
\usepackage{esint}
\usepackage[unicode=true, pdfusetitle, bookmarks=true, bookmarksnumbered=false, bookmarksopen=false, breaklinks=false, pdfborder={0 0 1}, backref=false, colorlinks=false]{hyperref}
\usepackage{graphicx}
\graphicspath{ {./images/} }
\usepackage{xcolor}
\usepackage[shortlabels]{enumitem}
\usepackage{makeidx}
\usepackage{amsfonts}
\usepackage{placeins}

\makeatletter
\newcommand{\ud}{\underline}

\makeatother

\begin{document}
\bibliographystyle{myunsrt}

\title{Bayesian Inference with Structured Signal: Static Replica Symmetry Breaking on the Nishimori Line in the Planted Spin Glass}

\author{Andrea Vincenzo Dell'Abate}
\affiliation{DI ENS, \'{E}cole Normale Supérieure, PSL Research University, CNRS, INRIA, Paris, France}

\author{Louise Budzynski}
\affiliation{DI ENS, \'{E}cole Normale Supérieure, PSL Research University, CNRS, INRIA, Paris, France}

\begin{abstract}
	A common assumption in theoretical models of Bayesian inference is that the signal has i.i.d. components. To study the effect of correlations in the signal prior, we consider a minimal model: the planted spin glass on random regular graphs, where the signal is sampled from an Ising model with coupling $\kappa$.
	Depending on the phase of the prior, we find that adding structure in the signal can either help or hinder inference.
	In the paramagnetic regime, correlations in the signal lower the reconstruction threshold, so that weaker signal strength is sufficient for recovery. In the ferromagnetic regime, the prior alone already enables partial recovery, and we identify the threshold above which the observations provide additional information.
	When the prior itself is in a replica symmetry breaking (RSB) phase, we detect a static RSB transition in the posterior under Nishimori conditions. This provides an example where a non-separable, correlated prior leads to static RSB in a Bayes-optimal inference problem. We discuss the consequences of this glassy phase for algorithmic performance, in particular for Belief Propagation.
\end{abstract}

\maketitle

\section{Introduction}
An inference problem consists of reconstructing a signal, law or pattern from partial and/or noisy observations (or data). In Bayesian inference, this task is achieved by weighting a prior belief on the signal with the likelihood of data. Inference problems appear ubiquitously in many scientific areas involving data analysis, such as signal processing \cite{Donoho_2009}, artificial intelligence \cite{pmlr-v202-cui23b}, computational biology \cite{Goldt_2023} and epidemiology \cite{Altarelli_2014, Braunstein_2023}. A central question in Bayesian inference is to assess under which conditions the information contained in the observations is enough to reconstruct the signal. Such an information-theoretic point of view should also be complemented by an algorithmic perspective: understanding what are the most efficient algorithms, and what is the best performance achievable computationally.

In many problems of interest, the signal to reconstruct and the observations are high-dimensional objects, making the theoretical analysis challenging. In such settings, tools from statistical physics, in particular the replica and cavity methods \cite{Parisi1987, Nishimori01}, led to a detailed description of the information-theoretical and algorithmic limits in many Bayesian inference problems \cite{Decelle_2011, Lesieur_2015}, predicting important properties, in particular computational-to-statistical gaps, i.e. regimes where reconstructing the signal is information-theoretically possible although no efficient algorithm exists. Many of these theoretical predictions were confirmed rigorously later on \cite{CojaOghlan_2018, Lelarge_2017}.

Although some recent efforts have focused on the theoretical study of structured datasets \cite{NEURIPS2019_Aubin, Goldt_2020, Loureiro_2022, Duranthon_2023, Ghio_2026}, most theoretical work on Bayesian inference model inputs as i.i.d. component-wise draws from some probability distribution. Despite providing valuable insights, these approaches are blind to the structure of real-world data, where one cannot assume that randomness enters in an uncorrelated way. Determining whether the presence of structure in the signal will be beneficial or detrimental to the inference process is a priori an open question and will depend on the particular problem to be considered. While it might make inference easier, as the knowledge of some structure in the signal can help to restrict the search to a smaller space, it might as well make the problem harder from an algorithmic viewpoint, as observed in \cite{Braunstein_2023, Braunstein_2025} when a Replica Symmetry Breaking (RSB) phenomenon prevents numerical methods (such as Simulated Annealing, or message-passing approaches) to provide a reasonable approximation of the posterior probability distribution of the signal given the data. This is particularly surprising, as it was conjectured that Bayes optimality implied the absence of glassy, static RSB phase at equilibrium \cite{Zdeborov__2016}. 

\begin{figure}
	\centering
	\includegraphics[width=\linewidth]{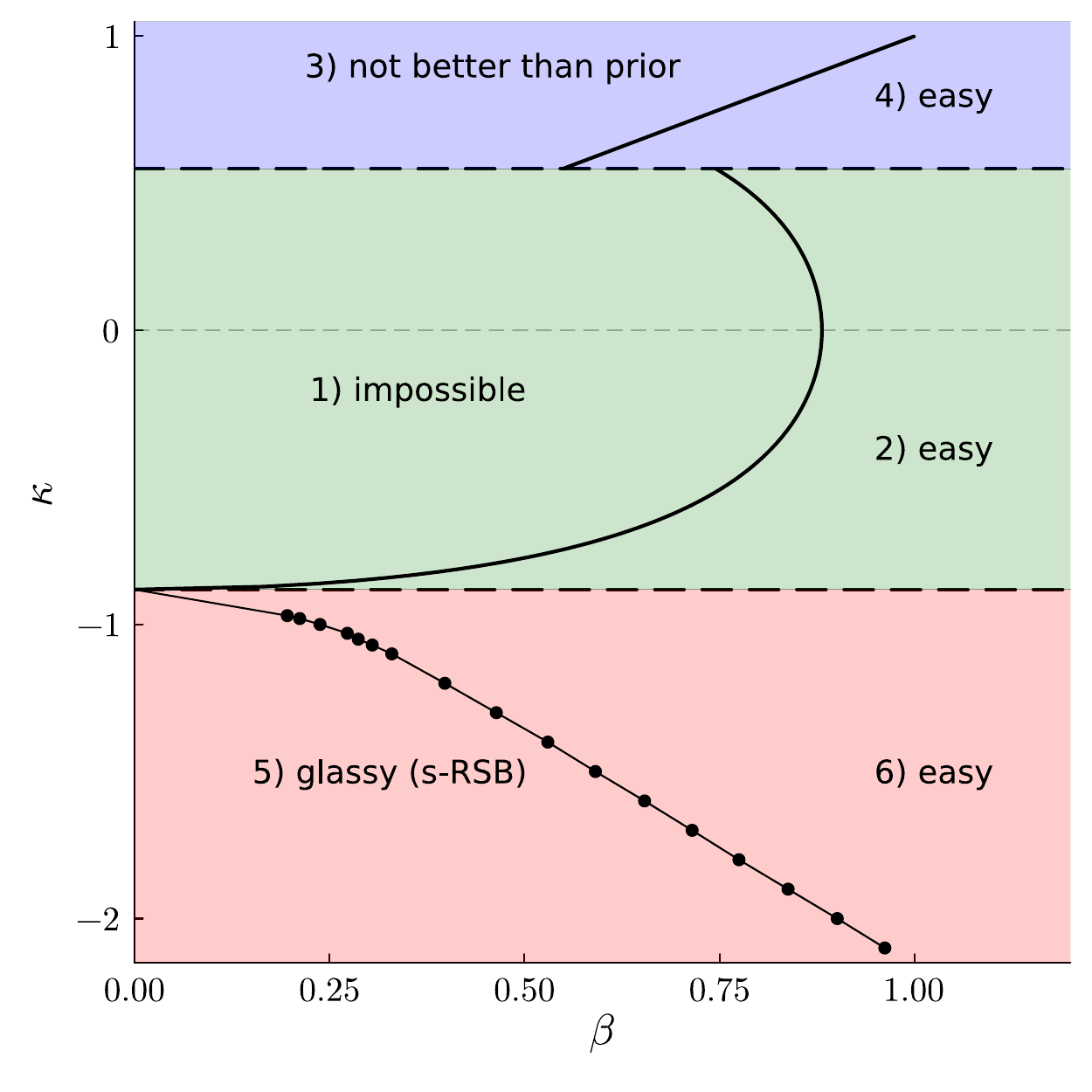}
	\caption{ Phase diagram of the inference problem obtained from the distributional cavity equations~(\ref{eq: posterior RS cavity equation}, \ref{eq:RS_cavity_eq_paramagnetic_prior}, \ref{eq: simplified 1RSB distributional cavity equation}), on random $3$-regular graphs. The background colors indicate the phase of the prior: paramagnetic for $\kappa\in[\kappa_{\rm rsb},\kappa_c]$ in green, ferromagnetic for $\kappa>\kappa_c$ in blue, and static RSB for $\kappa<\kappa_{\rm rsb}$ in red. In region 1, inference is information-theoretically impossible. In regions 2, 4 and 6, inference is possible and BP achieves the predicted optimal performance. In region 3, partial recovery is already possible from the prior alone, but the posterior does not improve over the trivial prior estimator. In region 5, the posterior is in a static RSB phase, within the $x=1$ cavity approach used here, the precise inference properties of this phase remain unresolved.}
	\label{fig:phase_diagram}
\end{figure}

In order to gain theoretical insights, we focus in this paper on the planted spin glass, also known as the Censored Block Model~\cite{Abbe_CBM_2013,saade_CBM_2015}, with a structured signal. In particular, we consider the signal, or planted configuration, to be sampled from the Boltzmann distribution of a homogeneous Ising model with inverse temperature $\kappa$, defined on a random regular graph. This model is minimal, analytically tractable, and its theoretical and computational limits are well understood in the case of unstructured ($\kappa=0$) signal \cite{Zdeborov__2016}. 

We study the information-theoretic and algorithmic properties of such model, focusing on the Bayes optimal setting.
We provide a quantitative analysis of the feasibility of the inference task, in the form of a phase diagram (see Figure~\ref{fig:phase_diagram}) depending on $\kappa$ (the parameter quantifying the signal's structure), and on the signal's strength $\beta$. The analysis is comprised of both a theoretical and a numerical part: in the former we study the properties of the posterior distribution in typical realizations of the inference problem on random regular graphs, focusing on the thermodynamic limit, by means of the cavity method. In the latter we perform a numerical resolution of the cavity equation, and compare its output with the results obtained by running Belief Propagation on finite size instances of the problem.

The prior probability, being the Boltzmann distribution of a homogeneous Ising model at inverse temperature $\kappa$ on random $d$-regular graphs, undergoes several phase transitions as $\kappa$ varies \cite{Zdeborova_Boettcher_2010, CojaOghlan_2020}. A ferromagnetic phase transition happens at $\kappa_c=\operatorname{atanh}({1}/{(d-1)})$, and a static Replica Symmetry Breaking (s-RSB) phase transition at $-\kappa_{\text{rsb}}=\operatorname{atanh}({1}/{\sqrt{d-1}})$. The phase diagram can therefore be divided into 3 regions, one for each phase of the prior. 

In the region where the prior is paramagnetic ($\kappa\in[\kappa_{\text{rsb}},\kappa_c]$), the phase diagram is divided into a region in which inference is easy (information-theoretically possible and with optimal performance achieved with Belief Propagation) and a region in which it is information-theoretically impossible. The critical value $\beta_c(\kappa)$ dividing the easy and impossible phases is lower than the one obtained with an unstructured prior $\beta_c(\kappa=0)$: the inference process is made easier as the introduction of the structure lowers the critical value $\beta_c$.

In the ferromagnetic phase ($\kappa>\kappa_c$), the prior becomes concentrated on the two states with non-zero magnetization. This effectively restricts the search space and allows an estimator based solely on the prior to achieve a non-trivial overlap with the signal. In this sense, adding structure to the signal renders the inference problem trivially easy, since partial recovery is already possible even without any observations. However, only above the critical threshold $\beta_{p}(\kappa)=\kappa$ can an estimator built from the posterior (hence incorporating the observations) attain an overlap larger than that achieved by the prior-based estimator. Therefore, it is only for $\beta>\beta_p$ that the observations provide information beyond what can already be inferred from the prior alone.

Finally, in the regime where the prior itself lies in a static RSB phase ($\kappa<\kappa_{\text{rsb}}$), we identify a critical threshold $\beta_{\text{rsb}}(\kappa)$ at which the posterior experiences a transition between a Replica Symmetric phase (for $\beta>\beta_{\text{rsb}}(\kappa)$) and a static RSB phase (for $\beta<\beta_{\text{rsb}}(\kappa)$). Since this transition occurs under Bayes-optimal conditions, it provides an example of a glassy, static RSB phase appearing on the Nishimori line. The mechanism behind this phenomenon stems from the strong and non-trivial correlations already present in the planted signal, which is sampled from a static RSB prior, unlike the standard setting where the signal components are independently distributed.
In the Replica Symmetric phase ($\beta>\beta_{\text{rsb}}(\kappa)$), the signal present in the observations is sufficiently strong for the posterior to decompose into two dominant states concentrated around the signal and its opposite. We find that this regime corresponds to an easy phase for the inference problem.
In the static RSB phase ($\beta<\beta_{\text{rsb}}(\kappa)$), our theoretical predictions are obtained by solving the cavity equation at Parisi parameter $x=1$, a choice that leads to significant simplifications and makes a numerical solution via population dynamics tractable. However, within this framework, the cavity method does not allow us to characterize the nature of the inference phase. Numerical simulations on finite-size systems nevertheless suggest that the structure of the signal in this regime hinders the inference task, as algorithms such as Belief Propagation struggle to converge for small values of $\beta$. 

The paper is organized as follows. In Section \ref{sec:set-up}, we introduce the inference problem under consideration. In Section \ref{sec:results}, we present our main results in the form of a phase diagram describing both the information-theoretically optimal performance and the algorithmic performance across the different regimes. A more detailed discussion is provided in Section~\ref{subsec:detailed_phase_diagram}, depending on the nature of the prior's phase (paramagnetic, ferromagnetic and static RSB).

\section{Set-up of the Problem}  
\label{sec:set-up}

\subsection{Definition of the model}
\subsubsection{The planted spin glass with structured prior}
We consider the planted spin glass model defined on a graph $G=(V,E)$. To each node $i\in V=\{1, \dots, N\}$ we assign a spin variable $s_i\in \{+1,-1\}$ in such a way that the planted configuration (the signal) $\ud s=\{s_1,\dots,s_N\}$ is sampled from the Boltzmann distribution of an Ising model with coupling constant $\kappa$ (serving as the prior distribution):
\begin{align}
\label{eq: prior}
P_{\kappa}(\ud{s})= \frac{1}{Z_\kappa(G)}\prod_{(i,j)\in E}e^{\kappa s_i s_j}\,.
\end{align}
On each edge $(i,j)\in E$, we introduce a coupling $J_{ij}\in\{+1,-1\}$, drawn independently conditional on the planted configuration $\ud s$. Its distribution is given by:
\begin{equation}
P(J_{ij}| s_i,s_j)=\rho \delta(J_{ij}-s_i s_j)+(1-\rho)\delta(J_{ij}+s_i s_j)\,.
\end{equation}
The collection of couplings $\ud J=\{J_{ij}\}_{\{i,j\}\in E}$ thus encodes noisy information about the alignment of spins $s_i$ and $s_j$: each edge reports the correct alignment $s_i s_j$ with probability $\rho$ and the opposite value with probability $1-\rho$.
Without loss of generality we can define the parameter $\beta$:
\begin{equation}
\beta = \dfrac{1}{2}\ln\left(\dfrac{\rho}{1-\rho}\right)\, ,
\end{equation}
which allows us to write the likelihood probability distribution as:
\begin{align}
\label{eq: likelihood}
P_{\beta}(\ud{J}|\ud{s})=\prod_{(i,j)\in E}\frac{e^{\beta J_{ij}s_i s_j}}{2\cosh(\beta)}\,.
\end{align}
The inference task is to recover the signal $\ud s$ from the noisy observations $\ud J$.
Applying Bayes' rule, we obtain the posterior distribution in the form of the Boltzmann distribution of a disordered Ising model:
\begin{equation}
\label{eq: posterior}
P(\ud s|\ud J) = \frac{1}{Z(G,\kappa,\beta,\ud J)}\prod_{(i,j)\in E}e^{(\kappa+\beta J_{ij})s_is_j} \,.
\end{equation}

\paragraph{Parameters of the model and known phase transitions.}
The parameter $\kappa$ quantifies the structure in the signal, while the parameter $\beta$ quantifies the signal's strength. At $\kappa=0$, the prior is uniform and one recovers the standard planted Ising model with unstructured signal~\cite{Zdeborov__2016}. Such model exhibits a phase transition at $\beta_c=\operatorname{atanh}({1}/\sqrt{d-1})$ on random $d$-regular graphs, separating an information-theoretically impossible phase at $\beta<\beta_c$, from an easy phase where known algorithms achieve information-theoretical optimal performance (which is strictly better than a random guess).

At $\beta=0$, the observations are pure noise, and the posterior probability distribution (\ref{eq: posterior}) equals the prior distribution (\ref{eq: prior}). When $\kappa>0$, the posterior corresponds to a ferromagnetic Ising model, for which a ferromagnetic phase transition~\cite{Can_2017_jstor} occurs at $\kappa_c = \operatorname{atanh}\left(1/(d-1)\right)$. When $\kappa<0$ (corresponding to the antiferromagnetic Ising model), a static RSB phase transition occurs at $\kappa_{\rm rsb}=-\operatorname{atanh}\left(1/\sqrt{d-1}\right)$~\cite{Zdeborova_Boettcher_2010, CojaOghlan_2020}.

\paragraph{Generalization of the inference problem.}
The above model can naturally be extended to more general settings. In particular, the observations $\ud J=\{J_{ij}\}_{\{i,j\}\in E}$ are collected on the edges of the graph $G=(V,E)$ defined for the prior (\ref{eq: prior}). A natural extension would be to collect the observations on a different graph $G'=(V',E')$ (e.g. Erd\H{o}s Rényi or random regular graphs).

The planted Ising model is equivalent to the Censored Block Model (CBM) in community detection, where two communities defined by their spin values have to be retrieved from censored edges observations. Another natural extension would be then to replace CBM observations by a Stochastic Block Model (SBM)~\cite{holland_SBM_1983, Decelle_2011}: an edge $(i,j)$ would be present in $E'$ with probability $p$ if $s_i=s_j$, and with probability $q$ if $s_i\neq s_j$. These natural extensions are left for future work.

\subsubsection{Relation with previous works}

Bayesian inference problems with non-trivial priors have previously been studied in several contexts. In spiked matrix models, signal sparsity has been incorporated through separable sparse priors, leading to efficient algorithms that achieve Bayes-optimal performance in some regimes~\cite{Deshpande_Montanari_2014}, while statistical-to-computational gaps emerge in other regimes~\cite{Lesieur_2015}. Structured sparse priors were also considered in~\cite{Jenatton_2010}, where exploiting additional correlations in the support structure was shown to improve performance compared to standard sparse PCA.

More recently, \cite{NEURIPS2019_Aubin} investigated the use of neural generative priors, replacing sparsity by a low-dimensional latent generative model. In the settings considered there, AMP achieves Bayes-optimal performance and no statistical-to-computational gap is observed. Neural priors have also been studied for the Stochastic Block Model~\cite{Duranthon_2023} and epidemic spreading~\cite{Ghio_2026}. While neural priors can significantly improve inference performance, first-order phase transitions and statistical-to-computational gaps may still arise in some regimes.

These works show that introducing structure in the prior can strongly affect the geometry of the inference problem. In particular, generative priors may improve inference performance by constraining the signal to a structured low-dimensional manifold, in contrast with unstructured uniform priors.
The approach taken in the present work is different: rather than introducing smooth low-dimensional structure, we study a prior whose degree of ruggedness can be tuned continuously through the homogeneous coupling constant $\kappa$.

\subsection{Bayesian inference and Bayes optimality}\label{subsec: bayesian inference and bayes optimality}
\subsubsection{Bayesian estimators}
As a measure of inference performance, we study in Section~\ref{sec:results} the overlap between the planted configuration $\ud s$ and an estimator $\hat{\ud \sigma}$:
\begin{equation}
    \operatorname{O}(\ud s, \hat{\ud \sigma})= \dfrac{1}{N}\sum_{i=1} s_i\hat \sigma_i \,.
\end{equation}

The best Bayesian estimate, the Mean Overlap, is obtained by averaging over the posterior distribution
\begin{equation}
    \operatorname{MO}(\hat{\ud\sigma})=\sum_{\ud \sigma}P_{\beta, \kappa}(\ud \sigma|\ud J)\dfrac{1}{N}\sum_{i=1}^{N}\sigma_i \hat\sigma_i\,.
\end{equation}

The \textit{Maximum Mean Overlap} (MMO) estimator is the one maximizing such quantity and is achieved for 
\begin{equation}\label{eq: MMO estimator}
    \hat{\sigma_i}^{\operatorname{MMO}}=\arg\max_{\sigma_i}P_{\beta,\kappa,i}(\sigma_i|\ud J)\,,
\end{equation}
where $P_{\beta,\kappa,i}(\sigma_i|\ud J)$ is the posterior marginal of node $i$. 
The overlap between the MMO estimator and the planted configuration provides a quantitative estimation of the accuracy of the estimator. 

Note that the posterior has a spin flip symmetry: a configuration $\ud \sigma$ and its flipped counter part $-\ud \sigma$ are statistically indistinguishable. Therefore, the posterior marginal $P_{\beta,\kappa,i}(\sigma_i|\ud J)$ is always uniform and the MMO estimator (\ref{eq: MMO estimator}) is uninformative. 

A valid definition of the estimator should take into account this spin-flip symmetry~\cite{Semerjian_2025}. In practice, when the posterior is in a paramagnetic phase the planted configuration is estimated by simply choosing each spin randomly with probability one half. Meanwhile, when the posterior is in the ferromagnetic or glassy phase, spontaneous symmetry breaking occurs (both under Belief Propagation iterations and in the iterative resolution of the cavity equations) and the system finds itself in one of the states with non-uniform marginal.

In the case of structured signal, the planted configuration undergoes a ferromagnetic transition for $\kappa>\kappa_c$. In this regime, the trivial estimator 
\begin{equation}\label{eq: trivial estimator}
	\hat\sigma_i^{\text{p}}=\arg\max_{\sigma_i}P_{\kappa,i}(\sigma_i)\,,
\end{equation}
maximizing the prior marginal of node $i$ achieves a non-vanishing overlap with the signal.
The relevant question is therefore whether the MMO estimator attains a larger overlap, or equivalently, whether the observations $\ud J$ provide information beyond that already contained in the prior.

\subsubsection{Bayesian optimality on the Nishimori line}

In the Bayes-optimal setting, the parameters used for inference match those of the generative model. In our case, this means that the prior parameter $\kappa$ and the channel parameter $\beta$ are known and used in the posterior distribution~(\ref{eq: posterior}). This setting corresponds to the Nishimori line, where the Nishimori identities hold~\cite{Nishimori01}, leading to an equality of the distribution $P(q)$ of the overlap between two configurations independently sampled from the posterior, and the distribution $P(q^*)$ of the overlap between the signal $\ud s$ and a configuration randomly sampled from the posterior.

In the case of an i.i.d. uniform prior, the gauge symmetry of the model allows one to map the planted overlap $q^*=O(\ud s,\ud \sigma)$ to the magnetization $m=\frac{1}{N}\sum_{i=1}^N\sigma_i$ of a spin glass model. Therefore, if the magnetization of the gauge-transformed model is self-averaging, the overlap distribution is also concentrated. This concentration is incompatible with a static RSB phase, whose signature would instead be a non-trivial overlap distribution~\cite{Zdeborov__2016}.

Note, however, that a dynamical RSB (d-RSB) phase may still be present. In such a phase, the Boltzmann measure splits into exponentially many clusters, each carrying a vanishing (exponentially small) weight. As a result, two independent samples typically belong to different clusters, so the overlap distribution $P(q)$ remains trivial. Dynamical RSB is therefore not visible from $P(q)$ or from finite-size correlations, but manifests itself through the clustered structure of the measure and the resulting slow dynamics, and can be detected by a positive complexity computed within the 1RSB cavity formalism.
 
For models with a separable prior, strong replica-symmetric properties are known to hold in the Bayes-optimal setting. In sparse models, the asymptotic decay of posterior two-point correlations was established in~\cite{Montanari_2007}. Such a decay is incompatible with a static spin-glass phase. 
More generally, under standard assumptions including concentration of the free energy, a strong form of replica symmetry, expressed as the concentration of all multioverlaps, was proved in~\cite{Barbier2022}. These results, however, rely on a separable prior, whereas in the present work the prior~\eqref{eq: prior} introduces correlations between the spin variables.

For the spiked Wigner model with generic structured priors, including generative neural priors, it was shown in~\cite{NEURIPS2019_Aubin}, see in particular Theorem~1, that the asymptotic mutual information is given by a replica-symmetric variational formula. In this sense, the spiked Wigner model does not exhibit a glassy static RSB phase at the level of the free entropy. This result is, however, specific to the spiked Wigner setting and does not by itself rule out static RSB phenomena in other inference models, such as the planted Ising model on sparse graphs studied here.

In epidemic inference, an RSB phase under Nishimori conditions was detected in~\cite{Braunstein_2025} through a stability analysis of the RS cavity solution. This provides another example where Bayes-optimal inference may exhibit RSB when the underlying variables are correlated: in that case, the infection times are constrained by the epidemic dynamics on the contact graph and therefore do not form a separable prior. The analysis identifies a continuous instability of the RS solution toward RSB, but does not solve the 1RSB cavity equations or characterize the nature of the RSB phase (static or dynamic).

Recent spin-glass results with correlated disorder suggest that the usual Nishimori-line argument against RSB is not robust to correlations. In particular, Nishimori showed that, for an Ising spin glass with a correlated disorder of a specific type, the magnetization distribution on the Nishimori line can become non-trivial and is related to the replica overlap distribution of the Edwards--Anderson model~\cite{Nishimori24}. Further works show that correlated disorder also leads to anomalous stability and chaos properties on the Nishimori line~\cite{Nishimori25,Nishimori_Ohzeki_Okuyama25}. These models are not inference problems, but they support the broader interpretation of our results: correlations in the generative structure, here induced by the structured prior, can allow a static RSB phase even under Nishimori conditions. 

\section{Results}
\label{sec:results}

\subsection{Phase diagram}
\label{subsec:phase_diagram}
In this section, we present the phase diagram of the planted spin-glass inference problem with a structured signal in the Bayes-optimal setting, see Figure~\ref{fig:phase_diagram} on random $3$-regular graphs.

The diagram can be divided in three regions, depending on the phase of the prior. The green region $\kappa\in[\kappa_\text{rsb}, \kappa_c]$ corresponds to the prior being in a paramagnetic phase. In this region we can distinguish an impossible phase, in which inference is information-theoretically impossible, and an easy phase, in which inference is information-theoretically possible and with optimal performance achieved by Belief Propagation. The two phases are separated by the critical value $\beta_c(\kappa)$ obtained via a stability study of the impossible solution (more details in Section~\ref{subsubsec:param_prior} and Appendix~\ref{app:stability_analysis}). 
Since $\beta_c(\kappa)<\beta_c(0)$ for all non-zero values of $\kappa$ in this regime, we conclude that the structure of the prior facilitates inference: compared with the unstructured case ($\kappa=0$), a lower signal strength is sufficient for reconstruction.
At the prior condensation transition, $\kappa=\kappa_{\rm rsb}$, we find $\beta_c=0$: arbitrarily weak observations are sufficient for reconstruction.
A natural interpretation is that the continuous nature of the transition makes the prior increasingly susceptible to small perturbations; as the correlation length, or susceptibility, diverges near $\kappa_{\rm rsb}$, even weak information from the observations can propagate through the system.

The blue region $\kappa> \kappa_c$ corresponds to the prior being in a ferromagnetic phase: the critical value $\beta_p(\kappa)=\kappa$ divides an easy phase and a ``not better than prior" phase, in which inference is feasible, but the MMO estimator~(\ref{eq: MMO estimator}) and the trivial estimator~(\ref{eq: trivial estimator}) achieve the same overlap with the signal. 

Finally, the red region $\kappa<\kappa_{\rm rsb}$ corresponds to a glassy, static RSB prior. 
In this regime, the critical threshold $\beta_\text{rsb}(\kappa)$ separates an easy phase, for $\beta>\beta_\text{rsb}(\kappa)$, in which the posterior is Replica Symmetric and inference is easy, and a glassy phase, for $\beta<\beta_\text{rsb}(\kappa)$, in which the posterior is in a static RSB phase. This transition is detected by the negative complexity of the thermodynamically dominant solution to the 1RSB cavity equations~(\ref{eq: simplified 1RSB distributional cavity equation}) at Parisi parameter $x=1$. 
We verified that, in the paramagnetic and ferromagnetic prior regimes, the 1RSB equation~\eqref{eq: simplified 1RSB distributional cavity equation} admits only the replica-symmetric solution, supporting the RS analysis in these phases. A static RSB posterior is detected only when the prior itself is static RSB and $\beta$ is sufficiently small.

In the 1RSB formalism, a static RSB, or condensed, phase corresponds to a decomposition of the Gibbs measure into a sub-exponential number of thermodynamically relevant clusters. Its correct description requires a Parisi parameter $x<1$, which selects the dominant clusters. Solving the full 1RSB equations at generic $x$ is substantially more involved, since it requires population dynamics over distributions of distributions. By contrast, the $x=1$ equations are much simpler and already allow us to detect the onset of condensation through the negativity of the complexity.
However, once a static RSB phase is detected, using $x=1$ no longer gives the correct thermodynamic observables, such as the overlaps and the corresponding MMO estimator. Therefore, the $x=1$ solution allows to locate the onset of the glassy phase but cannot fully characterize the associated inference phase. Numerical simulations nevertheless strongly suggest that this region corresponds to a hard phase, as discussed in Section~\ref{subsubsec:sRSB_prior}.

\subsection{A detailed picture of the phase diagram}
\label{subsec:detailed_phase_diagram}

\subsubsection{Paramagnetic prior}
\label{subsubsec:param_prior}
\begin{figure}[t]
	\centering
    \includegraphics[width=0.43\textwidth]{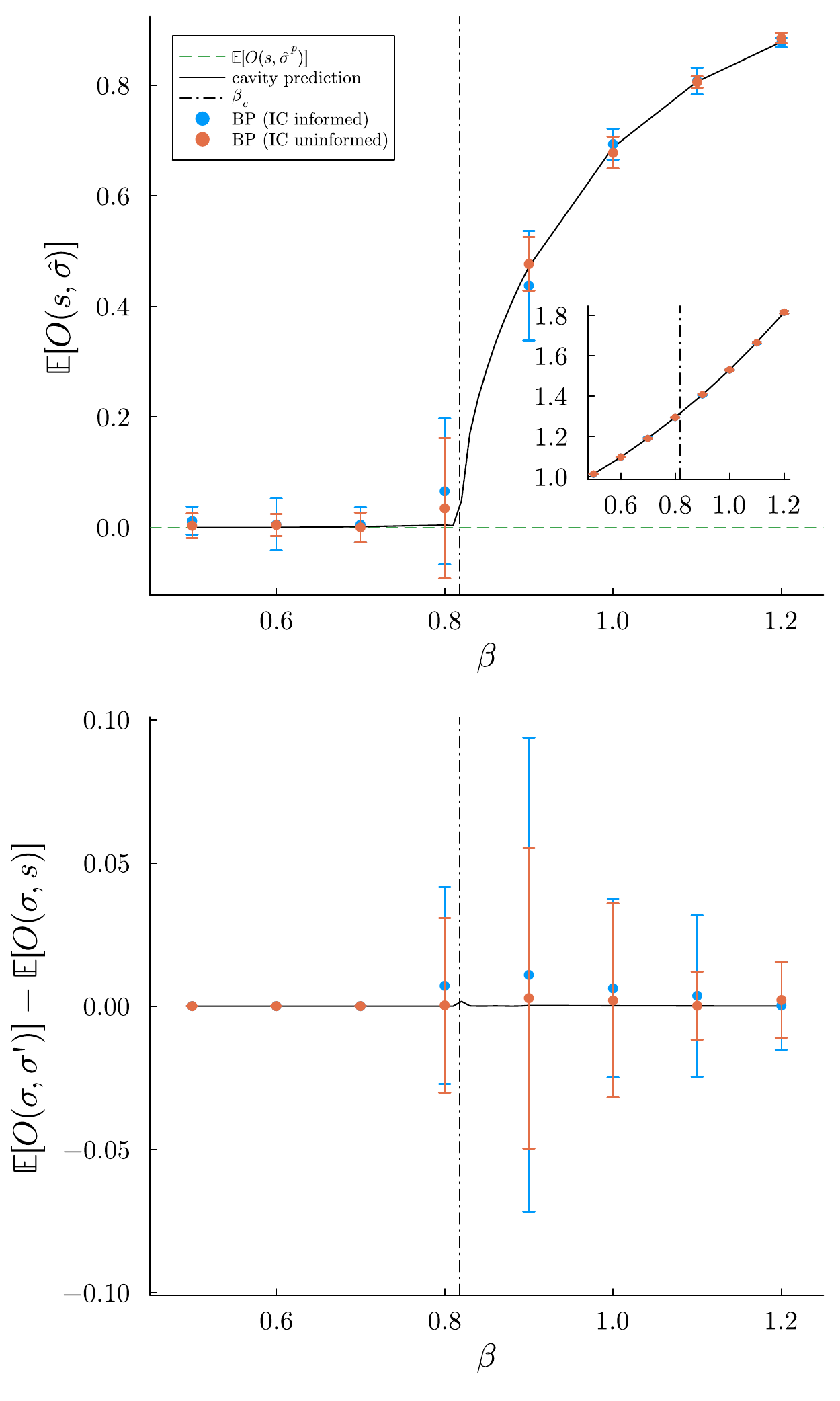}
	\caption{
	Transition for $\kappa=0.4$, from impossible to easy phase, occurring at $\beta_c=0.818$ (vertical dashed line). Top: Overlap between the MMO estimator~(\ref{eq: MMO estimator}) and the planted configuration $\ud s$ (the signal). Inset: Free-entropy of the posterior. Bottom: Nishimori identity, shown through the difference between the overlap $\mathbb{E}\left[\ud \sigma, \ud \sigma'\right]$ between two independent samples from the posterior, and the overlap $\mathbb{E}\left[\ud \sigma, \ud s\right]$ between the signal and a sample from the posterior. Black solid line: cavity predictions. Dots: performance of Belief Propagation on finite size instances $(N=10^4)$ with two initial conditions (see equation~(\ref{eq:BP_IC})) informed of the signal (blue dots), and uninformed (orange dots).}
	\label{fig:results_kappa0.4}
\end{figure}
For $\kappa\in [\kappa_{\text{rsb}},\kappa_c]$ the prior is in the paramagnetic phase, the planted configuration has vanishing magnetization and the trivial estimator~(\ref{eq: trivial estimator}) has zero overlap with the signal. 
The phase transition undergone by the posterior from the impossible phase to the easy phase is shown in Figure~\ref{fig:results_kappa0.4} for $\kappa=0.4$. The top panel displays the overlap between the signal and the MMO estimator, and the bottom panel shows that the Nishimori condition is satisfied. The inset displays the free-entropy of the posterior distribution (see equations~(\ref{eq:bethe_free_entropy_posterior}) and~(\ref{eq:free_entropy_RS}). Plots for different values of $\kappa$ can be found in Appendix~\ref{app:supplementary_plots}.

The results were obtained by solving the RS cavity equations~(\ref{eq:RS_cavity_eq_paramagnetic_prior}) numerically with population dynamics (see Appendix~\ref{app:RS_formalism} for the derivation).
The population of cavity messages was initialized as
\begin{equation}
\label{eq:IC_cavity_RS_param}
P(\nu^{(t=0)}|s)=\varepsilon\delta[\nu^{(t=0)}-\delta_{s,\cdot}]+(1-\varepsilon)\delta[\nu^{(t=0)}-1/2]\,,
\end{equation}
and equation (\ref{eq:RS_cavity_eq_paramagnetic_prior}) was recursively solved for $\textit{maxiter}=1000$ iterations, with population size $10^5$. For each value of the parameters, we used two initial conditions: one informed of the signal $\ud s$, with $\varepsilon=1$, and an uninformed one, with $\varepsilon=0.001$.

When several fixed points exist, the information-theoretic optimum is described by the thermodynamically dominant solution, i.e. the one with the largest free entropy. The algorithmic performance is probed by the fixed point reached from the uninformed initialization. 
We found a single fixed-point for all values of $\kappa,\beta$, thus discarding the presence of a hard phase.
In the impossible phase the overlap between the MMO estimator and the planted configuration is zero, while it is non-zero in the easy phase. 

The critical threshold $\beta_c(\kappa)$ separating the impossible and the easy phases, respectively for lower and higher values of $\beta$, can be obtained via a stability analysis of the impossible (paramagnetic) solution to equation~(\ref{eq:RS_cavity_eq_paramagnetic_prior}) (see Appendix \ref{app:stability_analysis}). In particular, for a fixed value of $\kappa$, the critical threshold $\beta_c(\kappa)$ must satisfy
\begin{equation}
     1=(d-1)\dfrac{\sum_{J}\tanh{(\kappa+\beta_c(\kappa) J)}\sinh{(\kappa+\beta_c(\kappa) J)}}{2\cosh{\kappa}\cosh{(\beta_c(\kappa)})}\,.
\end{equation}
We validate the cavity predictions on finite-size instances using Belief Propagation, whose results agree with population dynamics and therefore support the predictions of the cavity method.
Finite size instances of the inference problem were generated on random regular graphs of $N=10^4$ nodes, sampling a planted configuration using BP-guided decimation (see Appendix~\ref{app:sampling_planted_config}) and generating the observations conditionally on this signal. The messages were initialized as
\begin{equation}
\label{eq:BP_IC}
    \nu_{i\to j}(\sigma_i)=
    \begin{cases}
        \delta_{\sigma_i,s_i} & \text{w.p. }\varepsilon \\
        1/2 & \text{w.p. }1-\varepsilon
    \end{cases}\,,
\end{equation}
and BP iterations were run until all messages changed by less than $10^6$ between two consecutive iterations. Each simulation was performed twice, once for $\varepsilon=1$ and once for $\varepsilon=0.001$. For each set of parameters $(\kappa, \beta)$, the results were averaged over $10$ independent finite-size instances.

\subsubsection{Ferromagnetic prior}
\begin{figure}[t]
	\centering
	\includegraphics[width=0.43\textwidth]{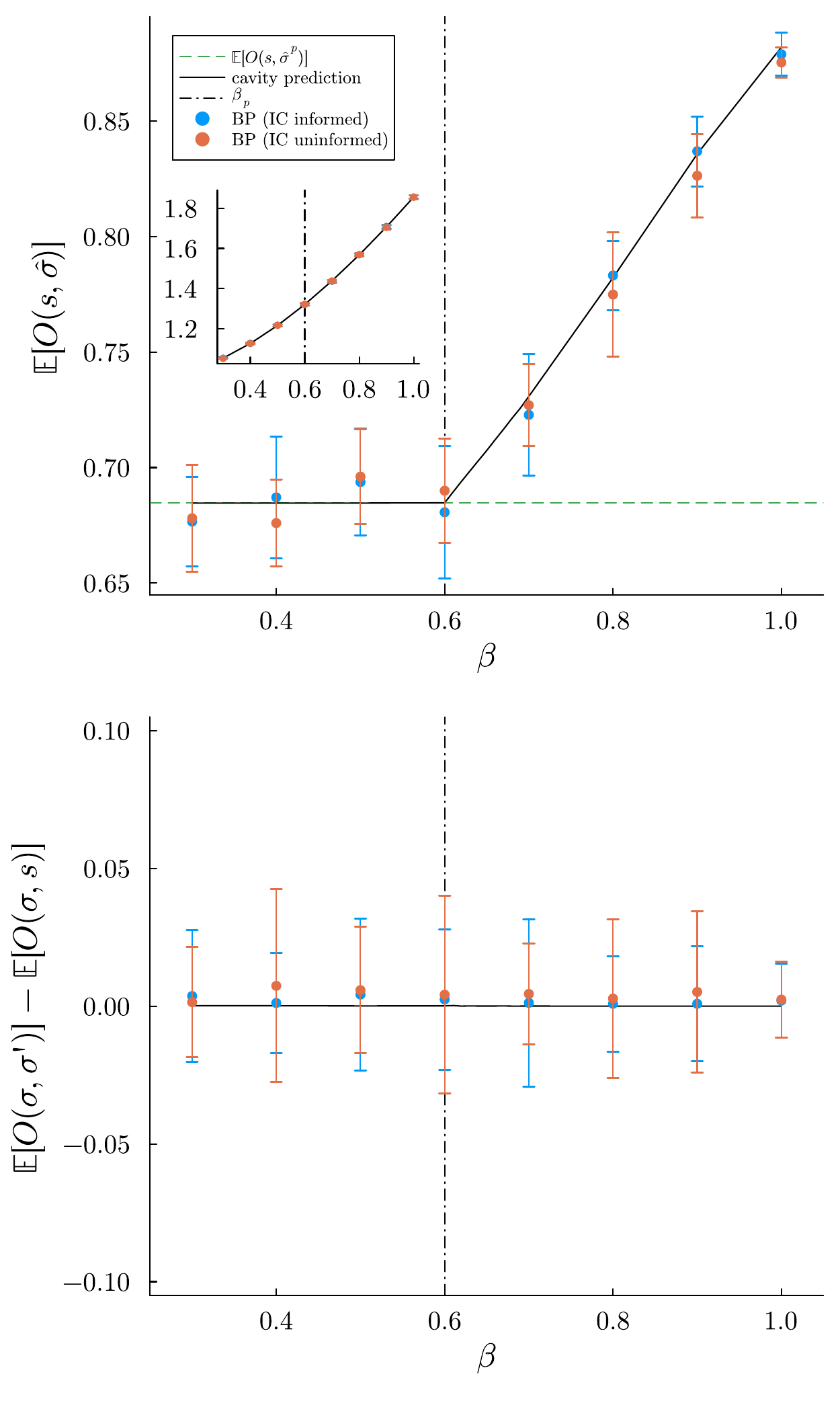}
	\caption{
	Transition for $\kappa=0.6$, from ``not better than prior'' to easy phase, occurring at $\beta_p=0.6$ (vertical dashed line). 
	Top: Overlap between the MMO estimator and the signal $\ud s$. For $\beta\leq \beta_p$, it equals the overlap achieved by the trivial estimator (green horizontal dashed line), and becomes larger for $\beta>\beta_p$.
	Inset: Free-entropy of the posterior.
	Bottom: Nishimori identity, shown through the difference between the overlap $\mathbb{E}\left[\ud \sigma, \ud \sigma'\right]$ between two independent samples from the posterior, and the overlap $\mathbb{E}\left[\ud \sigma, \ud s\right]$ between the signal and a sample from the posterior.
	Black solid line: cavity predictions. Dots: performance of Belief Propagation on finite size instances with two initial conditions (see equation~(\ref{eq:BP_IC_ferro})).
	}
	\label{fig:results_kappa0.6}
\end{figure}
When $\kappa > \kappa_c$, the prior is in the ferromagnetic phase and the trivial estimator~(\ref{eq: trivial estimator}) achieves non-zero overlap with the signal.
The phase transition is shown in Figure~\ref{fig:results_kappa0.6} for $\kappa=0.6$. It separates a ``not better than the prior'' phase, where the MMO estimator~(\ref{eq: MMO estimator}) achieves the same overlap as the trivial estimator, to an easy phase, where the MMO estimator achieves a strictly greater overlap. Plots for different values of $\kappa$ can be found in Appendix~\ref{app:supplementary_plots}.

The transition occurs at $\beta_p(\kappa)=\kappa$, which can easily be understood as follows.
For $\beta\leq \kappa$, all effective couplings in the posterior~\eqref{eq: posterior} are non-negative, since even a frustrated observation gives a coupling ($\kappa-\beta\geq0$). The posterior is therefore the Boltzmann measure of a ferromagnetic Ising model.

We observe that in this regime ($\beta\leq\kappa$) the system is in a ferromagnetic phase with two states of non-vanishing, opposite, magnetization, and the dynamics of BP iterations~(\ref{eq: posterior BP update equation}) and of the RS cavity equation (\ref{eq: posterior RS cavity equation}) fall and stay into one of the two states. Suppose that it falls into the state with positive magnetization: the local magnetization will be strictly positive for all nodes in the graph $\expval{\sigma_i}>0 \ \ \forall i\in V$,
which implies that 
\begin{equation}
    \hat{\sigma_i}^{\text{MMO}}=\arg\max_{\sigma_i}\mu_i (\sigma_i)=+1 \,,\qq{} \forall i\in V \,,
\end{equation}
i.e. the MMO estimator~(\ref{eq: MMO estimator}) and the trivial estimator (\ref{eq: trivial estimator}) are identical	: even with access to the posterior, one cannot improve over the information already contained in the prior. This is consistent with what happens at $\kappa\to +\infty$: the spins of the planted configuration are either all positive or all negative, and there is no additional information stored in the couplings $\ud J$ with respect to that already present in the prior. 

The results were obtained by solving the RS cavity equations specified for a ferromagnetic prior~(\ref{eq: posterior RS cavity equation}), with population dynamics. The difference is that, in the ferromagnetic phase, the uninformed initialization is not the uniform message but the prior BP fixed point. More precisely, the population of cavity messages was initialized as
\begin{equation}
\label{eq:BP_IC_ferro}
P(\nu^{(t=0)}|s,m^*)=\varepsilon\delta[\nu^{(t=0)}-\delta_{s,\cdot}]+(1-\varepsilon)\delta[\nu^{(t=0)}-m^*],
\end{equation}
with $m^*$ the BP fixed-point of the prior BP equations~(\ref{eq: prior BP update equation}) on a random regular graph.
For all values of ($\beta,\kappa$) considered, the informed and uninformed initializations converged to the same fixed point, thus discarding the presence of a hard phase within the RS cavity analysis.
For the comparison with finite-size BP, the same initialization scheme was used at the level of graph messages:
\begin{equation}
\nu_{i\to j}(\sigma_i)=
\begin{cases}
\delta_{\sigma_i,s_i} & \text{w.p. }\varepsilon \\
m^*(\sigma_i) & \text{w.p. }1-\varepsilon
\end{cases}\,,
\end{equation}
The BP results agree with the population-dynamics predictions, supporting the cavity analysis.

\subsubsection{Prior in a static RSB phase}
\label{subsubsec:sRSB_prior}
\begin{figure*}
    \centering
    \includegraphics[width=0.99\linewidth]{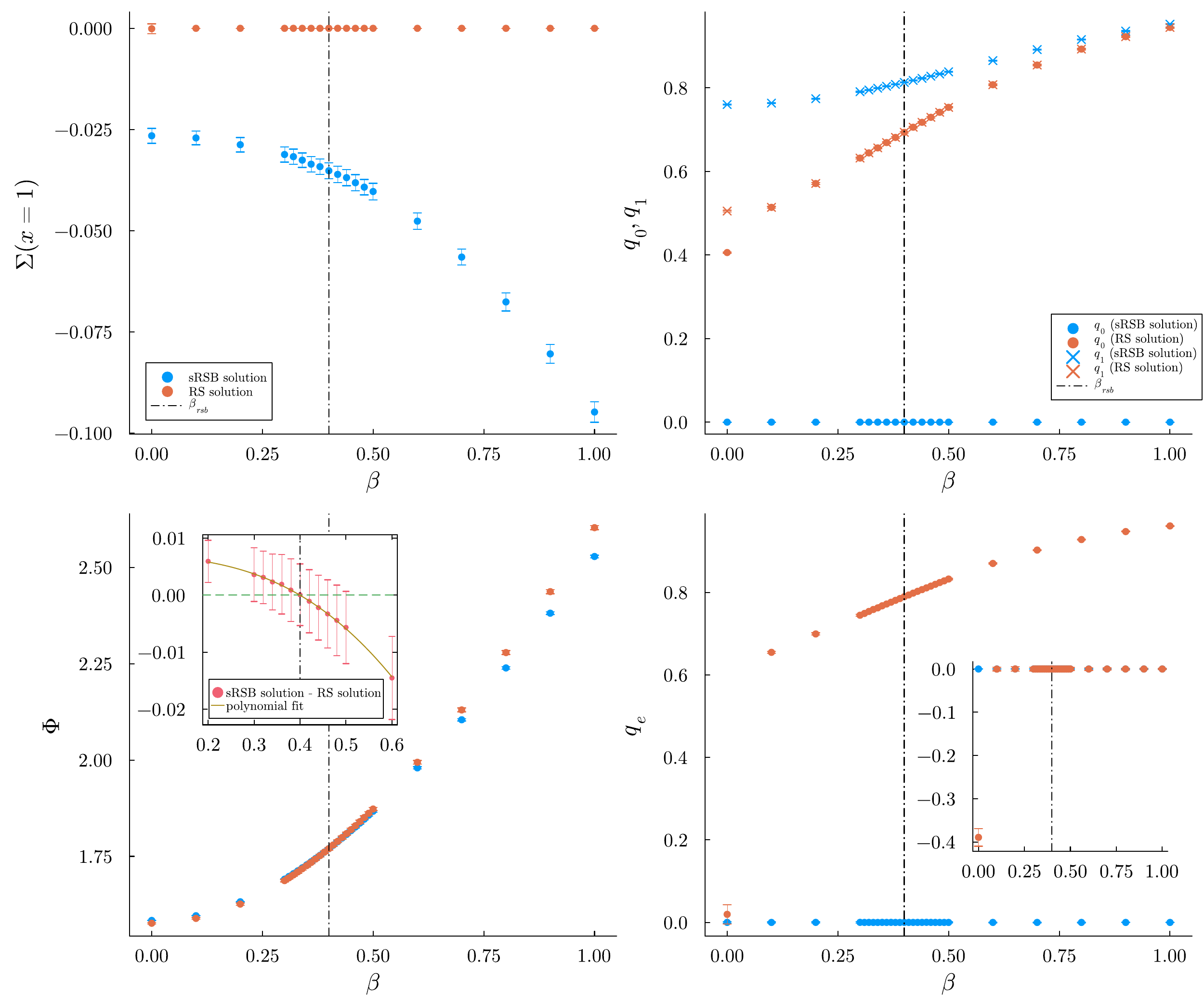}	
    \caption{
    Transition for $\kappa=-1.2$ from a static RSB phase to a easy RS phase, occurring at $\beta_{\rm rsb}=0.398$  (vertical dashed line).
    Top left: Complexity $\Sigma$ of the two solutions of~(\ref{eq: simplified 1RSB distributional cavity equation}).
    Top right: inter-state and intra-state overlap $q_0$ and $q_1$.
    The static RSB solution (blue markers) is characterized by $\Sigma<0$ and $q_0=0<q_1$. The RS solution (orange markers) has vanishing complexity and equal overlaps $q_0=q_1$, whenever it is thermodynamically dominant (the anomalous orange point at $\beta=0$ belongs to a subdominant continuation of the RS branch).
    Bottom left: Free-entropy of the two solutions. For $\beta<\beta_{\rm rsb}$, the s-RSB solution has larger free entropy, while for $\beta>\beta_{\rm rsb}$ the RS solution is dominant. The inset shows the free-entropy difference between the two solutions.
    Bottom right: Overlap $q_e=\mathbb{E}[O(\underline{s},\underline{\hat{\sigma}})]$. The static RSB solution has zero overlap with the signal, while the RS solution has non-zero overlap. 
    Inset: Nishimori identity, shown through the difference between the overlap $q_0$ between two configurations $\ud \sigma,\ud \sigma'$ sampled from the posterior, and a the overlap $q_p$ between a configuration sampled from the posterior $\ud \sigma$ and the planted configuration $\ud s$. This difference vanishes on the thermodynamically dominant branch, as expected on the Nishimori line.
	}
	\label{fig:results_kappa-1.2}
\end{figure*}
\begin{figure}[t]
	\includegraphics[width=0.45\textwidth]{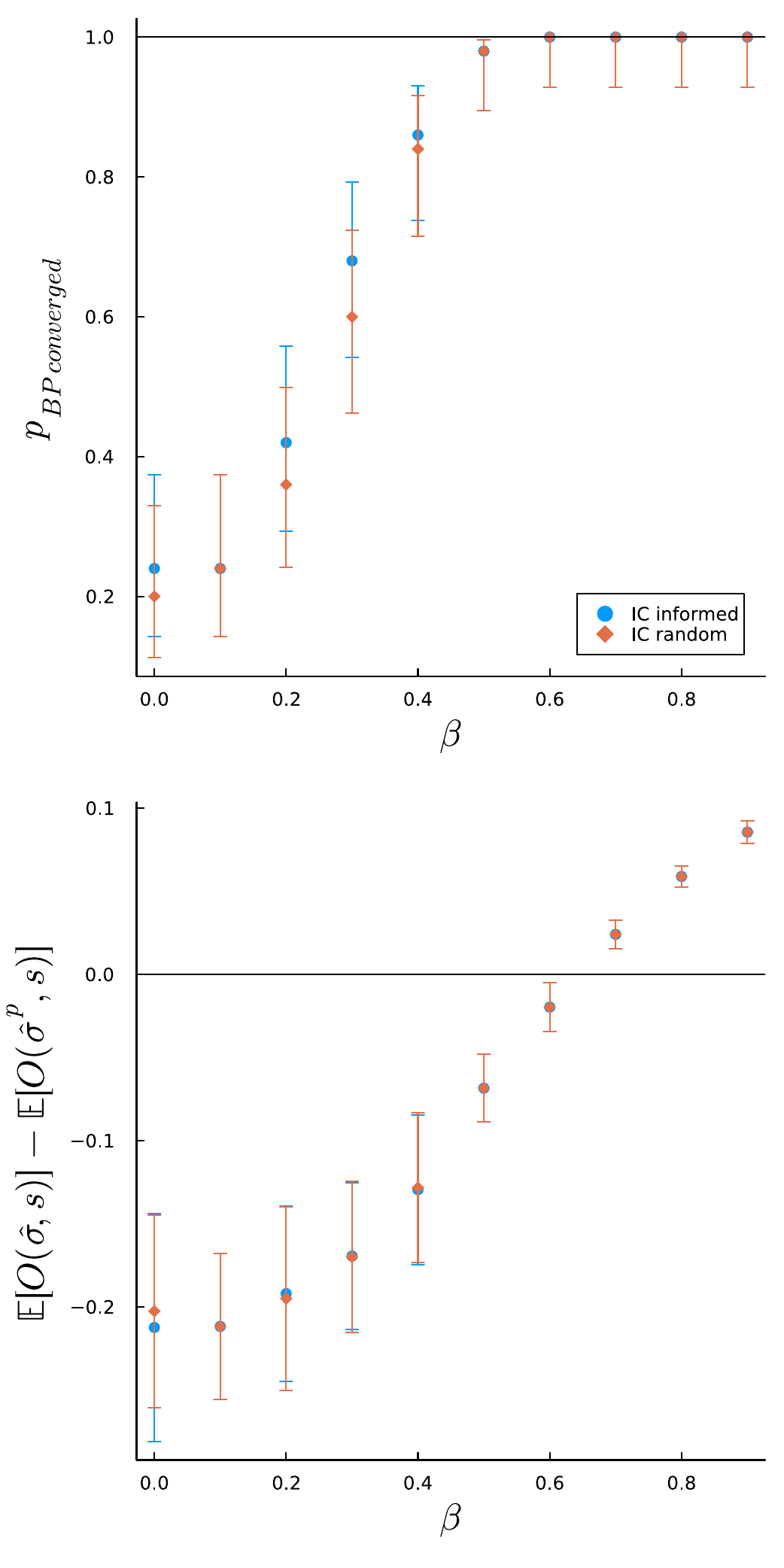}
	\caption{Performance of BP at $\kappa=-0.9<\kappa_{\rm rsb}\simeq -0.88$ on finite-size instances with $N=10^4$, averaged over $50$ samples.
	Top: convergence probability of BP.
	Bottom: difference between the overlap of the signal with the MMO estimator~\eqref{eq: MMO estimator} and with the prior-based estimator~\eqref{eq: trivial estimator}, averaged over converged instances.
	Blue markers correspond to the informed initialization, while orange markers correspond to the uninformed random initialization.
	}
	\label{fig:results_BP_kappa-0.9}
\end{figure}

When $\kappa<\kappa_{\text{rsb}}$, the prior lies in a static RSB phase. We identify a critical threshold $\beta_{\text{rsb}}(\kappa)$: for $\beta< \beta_{\rm rsb}$, the posterior is in a static RSB phase, while for $\beta>\beta_{\rm rsb}$ it is in a Replica Symmetric, easy phase. 

We solve the simplified 1RSB equation~\eqref{eq: simplified 1RSB distributional cavity equation} numerically using population dynamics (its derivation is described in Appendix~\ref{app:1RSB_formalism}).
Since the prior is itself in a static RSB phase, we first determine its Parisi parameter, denoted $x_0$. 
This step is relatively simple: the prior is an antiferromagnetic Ising model on a random regular graph, so its 1RSB cavity equation~\eqref{eq:prior_RSB_FP_eq} reduces to a single-population equation (see Appendix~\ref{subsubsec:estimating_x0} for details).
By contrast, solving the full 1RSB cavity equations for the posterior~\eqref{eq: 1RSB distributional cavity equation posterior rnd reg graphs} at a generic Parisi parameter $x$ is substantially more involved, since the unknowns are conditional distributions over distributions. 
We therefore rely on the simplifications that occur at $x=1$, leading to the tractable equations~\eqref{eq: simplified 1RSB distributional cavity equation}.

Once $x_0$ is fixed, the numerical resolution of~\eqref{eq: simplified 1RSB distributional cavity equation} proceeds in three stages. We first solve the 1RSB cavity equation of the prior~\eqref{eq:prior_RSB_FP_eq} (using population size $10^5$ and $1000$ iterations). The resulting population is then used to solve the RS cavity equation with an RSB prior~\eqref{eq: RS distributional cavity equation with RSB prior_RR}, which in turn provides the initialization for the simplified posterior 1RSB equation~\eqref{eq: simplified 1RSB distributional cavity equation}.
From the fixed point, we compute the free entropy $\Phi$, the complexity $\Sigma$, the inter-state and intra-state overlaps $q_0$ and $q_1$, the overlap between the MMO estimator~(\ref{eq: MMO estimator}) and the signal $q_e=\mathbb{E}\left[\ud s, \underline{\hat{\sigma}}\right]$, and the overlap $q_p=\mathbb{E}\left[\ud s, \underline{\sigma}\right]$ between the signal and a sample from the posterior (see Appendix \ref{app: RSB observables} for their explicit expression). 

We consider several initial conditions, defined in~(\ref{eq: first IC}--\ref{eq: sRSB initial condition}). 
All of them converge to the same RS solution, characterized by $q_0=q_1>0$ and $\Sigma=0$, except for the initialization~(\ref{eq: sRSB initial condition}) with $\varepsilon =0$, which converges instead to a static RSB solution with $q_0=0<q_1$ and $\Sigma(x=1)<0$.
The two solutions are shown in Figure~\ref{fig:results_kappa-1.2} for $\kappa=-1.2$.
We obtain the critical value $\beta_{\text{rsb}}(\kappa)$ by performing a fit (see the inset of Figure~\ref{fig:results_kappa-1.2}, bottom-left panel) of the difference between the free-entropies of the two solutions and finding the value of $\beta$ at which the difference changes sign.

\paragraph{\it Replica-Symmetric easy phase.} For $\beta>\beta_{\text{rsb}}(\kappa)$ , the thermodynamically dominant fixed point of~(\ref{eq: simplified 1RSB distributional cavity equation}) is replica-symmetric. 
Although the prior itself is in a static RSB phase, this is expected at large $\beta$: the observations are sufficiently informative for the posterior to concentrate around the planted configuration and its global flip, $\pm\ud s$.
The RS solution is characterized by zero complexity and coinciding inter-state and intra-state overlaps $q_0=q_1$, see the top panels of Figure~\ref{fig:results_kappa-1.2} in orange. Moreover, $q_0>0$, since the cavity dynamics selects one of the two planted states, as in the standard planted Ising model with $\kappa=0$.
Both informed and non-informed types of initial conditions lead to the same RS solution with non-zero overlap with the signal $q_e=\mathbb{E}[O(\underline{s},\underline{\hat{\sigma}})]>0$ (see Figure~\ref{fig:results_kappa-1.2}), indicating that this is an easy phase. This conclusion is further supported by the performance of the Belief Propagation algorithm on finite-size instances (see Figure~\ref{fig:results_BP_kappa-0.9}). 

\paragraph{\it Static RSB phase.}
As shown in the bottom-left panel of Figure~\ref{fig:results_kappa-1.2}, when $\beta$ is decreased below $\beta_{\rm rsb}(\kappa)$, the RS solution becomes subdominant and the static RSB solution takes over as the thermodynamically dominant one, with a larger free entropy. At $x=1$, this solution is characterized by a negative complexity and by distinct inter-state and intra-state overlaps, $q_0\neq q_1$ (see the top panels of Figure~\ref{fig:results_kappa-1.2} in blue). More precisely, the inter-state overlap is zero $q_0=0$ (enforced by the spin-flip symmetry), while the intra-state overlap is strictly positive $q_1>0$. 
As previously mentioned, the $x=1$ formalism allows us to detect the onset of the static RSB phase and to locate the transition $\beta_{\rm rsb}(\kappa)$. However, once inside this condensed phase, the physical solution should be described by the appropriate Parisi parameter $x<1$. The $x=1$ solution does not give the correct thermodynamic observables, in particular the overlaps, and therefore does not allow us to determine from the cavity computation alone whether this region is easy, hard, or impossible for inference.

\begin{figure}
    \centering
    \includegraphics[width=0.48\textwidth]{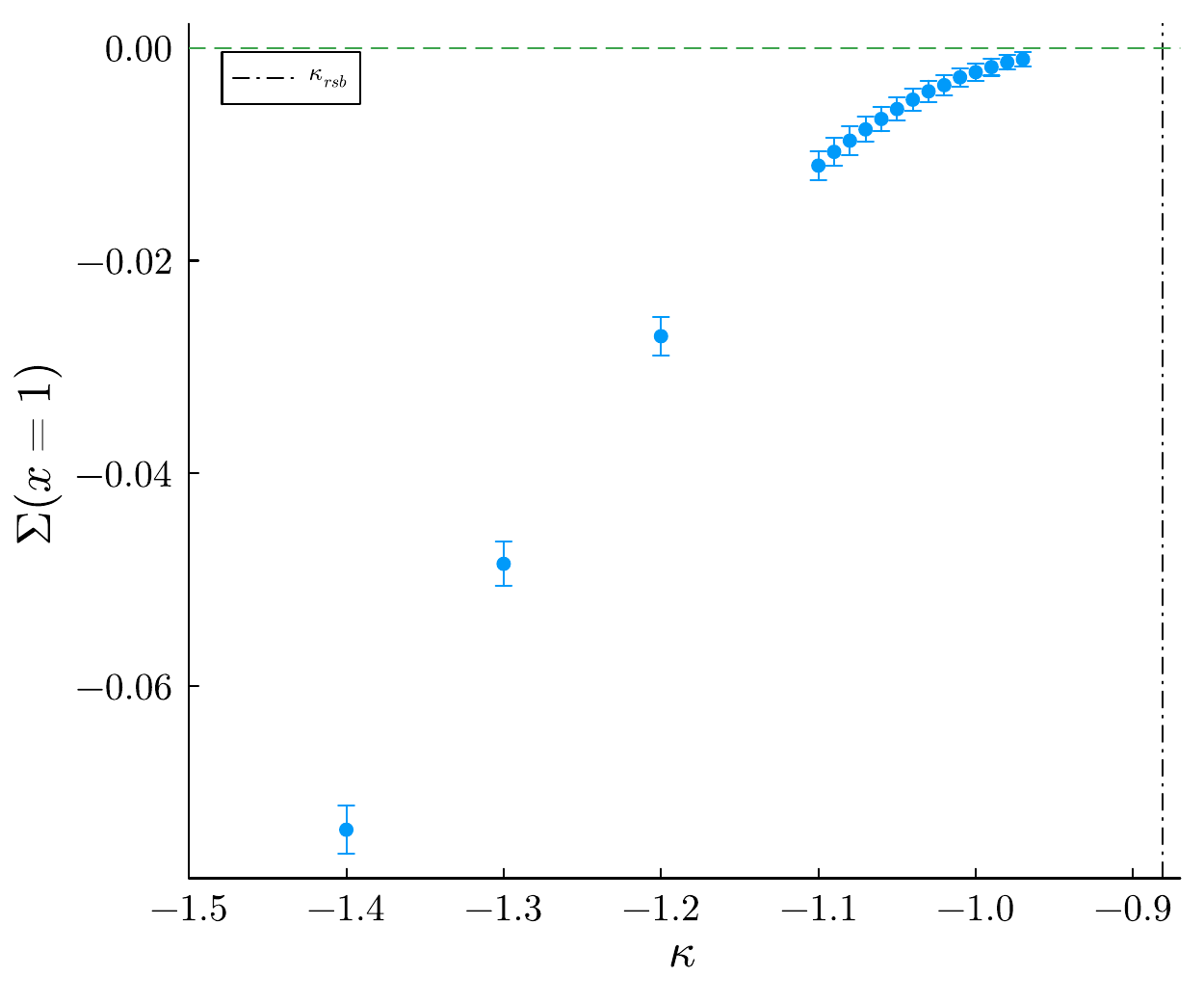}
    \caption{ Complexity $\Sigma$ of the static RSB solution of~(\ref{eq: simplified 1RSB distributional cavity equation}) at $\beta=0.1$ .}
    \label{fig:complexity_vs_kappa}
\end{figure}

Nevertheless, Figure~\ref{fig:complexity_vs_kappa} shows (for $\beta=0.1$) that, the complexity of the static RSB solution of~(\ref{eq: simplified 1RSB distributional cavity equation}) grows towards zero as $\kappa$ approaches $\kappa_{\rm rsb}$.
This suggests that the $x=1$ solution may still provide a reasonable approximation of the observables, for $\kappa$ values close to the critical threshold.
The $x=1$ cavity results suggest an unusual scenario. 
The thermodynamically dominant static RSB solution has zero overlap with the signal: $q_e=0$, which would normally indicate an information-theoretically impossible phase.
At the same time, some informed and uninformed initial conditions in~\eqref{eq: first IC}--\eqref{eq: sRSB initial condition} converge to subdominant solutions with $q_e>0$, which can be interpreted as metastable informative states.
This suggests that recovery might be possible by converging to such states, despite the dominant ($x=1$) solution being uninformative. 
This does not correspond to the standard easy, hard, or impossible scenarios~\cite{Ricci_Tersenghi_2019}, and should therefore be interpreted with caution.

Figure~\ref{fig:results_BP_kappa-0.9} shows the BP results for $\kappa=-0.9<\kappa_{\rm rsb}\simeq-0.88$ on finite-size instances of random $3$ regular graphs with $N=10^4$. 
As $\beta$ decreases, the performance of BP deteriorates: the probability of convergence drops, and the difference $\mathbb{E}\left[O(\underline{s},\underline{\hat{\sigma}}^{\rm MMO})\right]-\mathbb{E}\left[O(\underline{s},\underline{\hat{\sigma}}^{\rm p})\right]$ becomes negative. 
Thus, in this low-$\beta$ regime, BP run on the posterior performs worse than the prior-based estimator.
Cavity predictions are not shown in Figure~\ref{fig:results_BP_kappa-0.9}, because estimating $x_0$, and hence $\beta_{\rm rsb}$, is numerically unreliable so close to $\kappa_{\rm rsb}$, where the prior complexity is nearly zero (see Appendix~\ref{subsubsec:estimating_x0} for more details on the determination of $x_0$).

Again, these finite-size simulations should be interpreted with caution, since sampling from the prior is challenging in the static RSB phase. We generate planted configurations using BP-guided decimation, see Appendix~\ref{app:sampling_planted_config}, but this method is expected to give accurate marginals only in the RS phase. Its reliability is therefore not guaranteed once the prior enters the RSB regime.
Nevertheless, because the transition to the static RSB phase is continuous, we expect the effects of clustering to appear gradually as $\kappa$ approaches $\kappa_{\rm rsb}$. BP-guided decimation may therefore still provide reasonable approximate samples close to the transition. As a consistency check, we verified that the sampled planted configurations have an energy in agreement with the cavity predictions.

Since these results are obtained on the Nishimori line, with the inference parameter $\kappa_{\rm inf}$ matched to the value $\kappa$ used to generate the signal, one may ask whether using a mismatched prior could improve BP. We tested the choice $\kappa_{\rm inf}=0$, corresponding to an unstructured prior. Although BP then converges on all tested instances and for all values of $\beta$, its overlap is systematically smaller than that of the prior-based estimator, and worse than in the matched case $\kappa_{\rm inf}=\kappa$ (data not shown).

Overall, the BP results suggest a hard inference regime: for sufficiently small $\beta$, BP convergence deteriorates and the posterior-based estimator no longer improves over the prior-based one. 
As shown in the phase diagram of Figure~\ref{fig:phase_diagram}, for sufficiently negative values of the structure parameter $\kappa$, the static RSB transition occurs at $\beta_{\rm rsb}(\kappa)>\beta_c(\kappa=0)$, above the critical threshold of the unstructured prior. 
In this sense, adding structure to the signal makes inference harder in this regime, by reducing the range of $\beta$ values for which inference is algorithmically easy.

\section{Discussion and Perspectives}
In this work, we studied a minimal and analytically tractable Bayesian inference problem with a structured signal: the planted spin glass on random regular graphs.
In many high-dimensional Bayesian inference models, the signal is assumed to have i.i.d. components, corresponding to a separable prior, while the observations are independent conditionally on the signal. 
Under such hypothesis, strong replica-symmetric properties are expected in the Bayes-optimal setting, and in some cases this can be proved rigorously~\cite{Montanari_2007,Barbier2022}.
Recent works, however, have shown that going beyond these standard assumptions can significantly modify the phase diagram, both from an information-theoretic and an algorithmic point of view~\cite{NEURIPS2019_Aubin,Duranthon_2023,Braunstein_2025,Ghio_2026}.
Here, we go beyond one of these two standards assumptions by considering a non-separable prior: the signal is sampled from an Ising model with coupling $\kappa$, which introduces correlations between its components.

Our results show that this structure can have qualitatively different effects depending on the phase of the prior.
When the prior is paramagnetic, the structure facilitates inference: the critical value $\beta_c(\kappa)$ is lower than in the unstructured case $\kappa=0$, therefore a weaker signal in the observations is sufficient for reconstruction.
When the prior is ferromagnetic, the prior itself already allows partial recovery, since the signal is concentrated around two magnetized states. In this regime, the relevant threshold is no longer the onset of reconstruction, but the value $\beta_p(\kappa)=\kappa$ above which the posterior estimator improves over the prior-based one.

The most striking regime occurs when the prior itself is in a static RSB phase. 
In this case, we find that the posterior undergoes a transition, under Bayes-optimal conditions, between an easy replica-symmetric phase at large $\beta$ and a static RSB phase at low $\beta$. 
This provides an example of a static RSB phase on the Nishimori line in a Bayesian inference problem. 
This phenomenon could be explained by the correlations present in the signal: unlike in separable-prior models, the planted configuration already carries a non-trivial glassy structure, which can persist in the posterior when the observations are not informative enough to select the planted state.
Our result is part of a recent line of work showing that RSB can arise under Nishimori conditions, both in Bayesian inference~\cite{Braunstein_2023,Braunstein_2025} and in spin-glass models with correlated disorder~\cite{Nishimori24,Nishimori25,Nishimori_Ohzeki_Okuyama25}.

The analysis of this glassy regime relies on the $x=1$ 1RSB cavity equations. This approach makes the numerical study tractable and allows us to detect the onset of the static RSB phase through the negativity of the complexity.
However, once the condensed phase is reached, the correct thermodynamic description should involve a Parisi parameter $x<1$. 
As a consequence, the $x=1$ solution does not provide the correct thermodynamic observables inside the static RSB phase, and the nature of the  inference phase this region remains unresolved within the present cavity computation. 
Finite-size BP simulations nevertheless suggest that this region behaves as a hard phase, with deteriorating convergence and posterior-based estimates that may fail to improve over the prior-based estimator.

Several directions remain open. A first natural step would be to solve the full 1RSB cavity equations at the appropriate Parisi parameter $x<1$, in order to compute the correct overlaps and fully characterize the inference phase inside the condensed region. It would also be interesting to study variants of the model, for instance by replacing the planted Ising model, which is related to the censored block model in community detection, with the stochastic block model.
Another direction currently investigated is the one of dense models with structured signals. While replica symmetry is known to hold for the spiked Wigner model with generic priors~\cite{NEURIPS2019_Aubin}, the situation for spiked tensor models remains an interesting question. Similarly, one could consider higher-order interactions, for example Bayesian inference problems on hyper-graphs such as planted constraint satisfaction problems. In the unstructured case, such higher-order models often exhibit hard phases; understanding how these phases are modified by a structured signal would be interesting.
Finally, one could go beyond the assumption of conditionally independent observations. This is especially relevant in teacher--student supervised learning, where it amounts to relaxing the independence assumption between samples. More broadly, our results support the idea that moving beyond the simplifying assumptions traditionally used in statistical physics and high-dimensional inference can qualitatively change the phase diagram. 
In particular, correlated priors provide a natural setting in which the standard arguments for replica symmetry in Bayes-optimal settings may no longer apply. 
It would be interesting to identify more general conditions under which static RSB can occur in Bayes-optimal settings.

\acknowledgments
We thank Alfredo Braunstein, Odilon Duranthon and Guilhem Semerjian for insightful discussions. 
This work has received support under the 2025 PSL Young Researcher Starting Grant operated by Université PSL (grant agreement no. 2025-392).
A public GitHub repository containing an implementation of the algorithm and notebooks is available at~\cite{github_ref}.

\bibliography{biblio.bib}

\appendix

\section{Belief Propagation}

\subsection{Belief-Propagation equations}
For each edge $(i,j)\in E$ we introduce the BP messages $\nu_{i\to j},\nu_{j\to i}$ as the marginal probability distributions of $\sigma_i$ and $\sigma_j$ in the amputated graph $G=(V,E\setminus\qty{(i,j)})$.
The BP messages obey the following set of equations defined on each directed edge $i\to j$: 
\begin{equation}\label{eq: posterior BP update equation}
    \nu_{i\to j}(\sigma_i)=\dfrac{1}{z_{i\to j}} \prod_{k\in \partial i \setminus j}\sum_{\sigma_k} e^{(\kappa+\beta J_{ik})\sigma_i\sigma_k}\nu_{k \to i}(\sigma_k)\,,
\end{equation}
where $z_{i\to j}$ is the normalization factor. The marginal probability of $\sigma_i$ can be computed from the set of incoming messages $\qty{\nu_{k\to i}}_{k\in\partial i}$
\begin{equation}\label{eq: posterior BP marginal equation}
    \mu_{i}(\sigma_i)=\dfrac{1}{z_{i}} \prod_{k\in \partial i}\sum_{\sigma_k} e^{(\kappa+\beta J_{ik})\sigma_i\sigma_k}\nu_{k \to i}(\sigma_k)\,,
\end{equation}
where $z_{i}$ is the normalization factor and $\partial i$ is the set of neighbors of vertex $i$.
The free entropy of the posterior can be estimated through the Bethe free entropy:
\begin{equation}\label{eq:bethe_free_entropy_posterior}
    \Phi=\dfrac{1}{N}\sum_{i=1}^N\ln z_i -\dfrac{1}{N}\sum_{i=1}^{N}\ln z_{ij}\,,
\end{equation}
where 
\begin{equation}
    z_{ij}=\sum_{\sigma_i,\sigma_j} e^{(\kappa+\beta J_{ij})\sigma_i\sigma_j}\nu_{i \to j}(\sigma_i)\nu_{j \to i}(\sigma_j)\,.
\end{equation}
BP messages, marginal probabilities and Bethe free-entropy for the prior can be obtained by setting $\beta=0$ in the above formulas:
\begin{align}
    m_{i\to j}(s_i)&=\dfrac{1}{z^{0}_{i\to j}} \prod_{k\in \partial i \setminus j}\sum_{s_k} e^{\kappa s_i s_k}m_{k \to i}(s_k)\,,\label{eq: prior BP update equation}\\
    b_{i}(s_i)&=\dfrac{1}{z_{i}^0} \prod_{k\in \partial i}\sum_{s_k} e^{\kappa s_is_k}m_{k \to i}(s_k)\,,\label{eq: prior BP marginal}\\
    \Phi^0&=\dfrac{1}{N}\sum_{i=1}^N\ln z^0_i -\dfrac{1}{N}\sum_{i=1}^{N}\ln z^0_{ij}\,,\\
    z^0_{ij}&=\sum_{s_i,s_j} e^{\kappa s_is_j}m_{i \to j}(s_i)m_{j \to i}(s_j)\,.
\end{align}

\subsection{Sampling the planted configuration}\label{app:sampling_planted_config}
In the finite-size experiments, the planted configuration $\ud s$ is sampled from the prior distribution~\eqref{eq: prior}. This could be done using Monte Carlo sampling or Belief Propagation algorithms. In practice, we use a BP-guided decimation procedure~\cite{Montanari_2019}, which is efficient and gives accurate samples in the regimes where BP provides reliable estimates of the marginals.

On a random regular graph, BP marginals are only approximate, but the graph is locally tree-like, and this approximation is expected to be accurate in the RS regime. We proceed as follows. 
We first choose a node $i$ uniformly at random and sample the spin $s_i$ from the BP estimate of the prior marginal~(\ref{eq: prior BP marginal}). 
We then define the two sets 
\begin{equation}
     S=\qty{i\in V: s_i \text{ has been fixed}} \qq{and} R=V\setminus S\,,
\end{equation}
At each subsequent step, we choose a vertex $j\in R$ uniformly at random and sample $s_j$ from the BP estimate of the marginal of the conditional prior $P_\kappa(\underline{s}_R\mid \underline{s}_S)$ (with $\ud s_A =\qty{s_k}_{k\in A}$ for $A=R,S$) where the spins in $S$ are fixed to their previously sampled values.
To estimate this marginal probability we run BP again, adding for each $i\in S$ a factor node enforcing the value $s_i$. We add $j$ to $S$ and repeat the procedure until all spins have been fixed ($S=V$).

This procedure is exact if one is able to exactly compute the marginal probabilities. In that case the procedure is equivalent to decomposing the prior (\ref{eq: prior}) using the chain rule. In BP-guided decimation we approximate these marginals using Belief Propagation.
As discussed in the main text, this approximation is less controlled when the prior is in a static RSB phase, where BP marginals are no longer guaranteed to be accurate.

Finally, when $G$ is a tree, the sequential sampling procedure simplifies. 
Starting from a vertex $i$, we first sample $s_i$ from the BP marginal~\eqref{eq: prior BP marginal}, and then recursively sample the spins of its descendants down to the leaves. 
Since BP marginals are exact on trees, this gives an exact sampler. If $j$ is a child of $i$, the conditional distribution used in the recursion is
\begin{align} 
	P(s_j|s_i, m_{j\to i})=\dfrac{e^{\kappa s_i s_j}m_{j\to i}(s_j)}{\sum_{s_j}e^{\kappa s_i s_j}m_{j\to i}(s_j)} \,.
\end{align}
This tree sampling procedure is the one underlying the derivation of the cavity equations in the following sections.

\section{Cavity method under the Replica Symmetric Ansatz}
\label{app:RS_formalism}
Deriving distributional cavity equations is non-trivial when the disorder variables are not i.i.d., as is the case for the posterior couplings $\ud J=\{J_{ij}\}_{\{i,j\}\in E}$ in our model. 
In this section and the following one, we derive the cavity equations under the RS and the 1RSB ansatz. 
Note that similar cavity equations at generic $K$-step of RSB were already derived in~\cite{Semerjian2008}.

Consider the following thought experiment. Draw a graph $G$ from the random $d$-regular graph ensemble, sample the planted configuration $\ud s$ from the prior probability (\ref{eq: prior}) and the couplings $\ud J$ from the likelihood probability (\ref{eq: likelihood}).
After running Belief Propagation to convergence and obtaining a fixed point, choose an edge $i \to j$ uniformly at random. 
The resulting message $\nu_{i\to j}$ associated to this edge is a random variable with probability distribution $P(\nu_{i \to j})$. 

The replica-symmetric cavity method assumes a fast decay of the correlations between distant variables in the measure~(\ref{eq: posterior}). 
Under this assumption, in the thermodynamic limit, a solution to the BP equations exists, and this solution approximates well the marginals of the model.
This entails that the message $\nu_{i\to j}$ converges in distribution to a random variable $\nu$, whose probability distribution $P(\nu)$ satisfies a fixed point equation
\begin{equation}
    P(\nu)=\mathcal{F}^{\text{RS}}(P(\nu))\,,
\end{equation}
called RS distributional cavity equation.

\subsection{Cavity equations for the prior probability}\label{app:RS_cavity_eq_prior}
In this section we start by illustrating how to obtain the distributional cavity equation for the prior probability (\ref{eq: prior}) (for which the standard derivation applies) by retracing the steps shown in \cite{Mezard_2009}.

Let $m_{i_0 \to j_0}^{(t)}$ be the message sent by the BP algorithm at iteration $t$ along the edge $(i_0,j_0)$. Let us assume that the initial messages $\ud m^{(0)}$ are i.i.d. random variables with distributions independent of $N$. 

Let us introduce the directed neighborhood of radius $t$ of a directed edge $i_0\to j_0$, denoted by $B_{i_0 \to  j_0}^{(t)}$. This is the subgraph of $G$ made up of all the nodes which can be reached from $i_0$ by a non-reversing path of length at most $t$, when edge $(i_0,j_0)$ is removed. 

Because $G$ is sampled from the random $d$-regular graph ensemble, in the thermodynamic limit $N \to \infty$ and for fixed $t$, the neighborhood $B_{i_0\to j_0}^{(t)}$ converges in distribution to $T_{i_0 \to j_0}^{(t)}$, a full $(d-1)$-ary regular tree of depth $t$ with root edge $i_0\to j_0$. All the nodes of the tree, except for the leaves, have $d-1$ children.

The message $m_{i_0 \to j_0}^{(t)}$ is a function of the graph $G$ and the initial conditions $\ud m^{(0)}$. However the dependence on $G$ only occurs on through the directed neighborhood $B_{i_0 \to j_0}^{(t)}$, while the dependence on the initial condition is only through the messages 
\begin{equation}
    \ud m^{(0)}_{\partial B_{i_0 \to j_0}^{(t)}}=\qty{m^{(0)}_{k\to p(k)}: k\in \partial B_{i_0\to j_0}^{(t)}}\,,
\end{equation}
where $\partial B_{i_0\to j_0}^{(t)}$ is the set of vertices of $B_{i_0\to j_0}^{(t)}$ exactly at distance $t$ from $i_0$, and $p(k)$ is the parent of $k$ in the directed neighborhood.

Consider the case in which the neighborhood of $(i_0,j_0)$ is really a regular full $(d-1)$-ary random tree $T_{i_0 \to j_0}{(t)}$. We define $m^{(t)}$ as the message passed through the root edge of this tree after $t$ BP iterations. Since $B_{i_0 \to j_0}^{(t)}$ converges in distribution to the tree $T_{i_0 \to j_0}^{(t)}$, we find that $m^{(t)}_{i_0 \to j_0}\xrightarrow{d}m^{(t)}$ as $N \to \infty$. Let us now consider an edge $k\to l$ at a distance $r$ from the root and directed towards it. The directed sub-tree rooted in $k\to l$ will be a full $(d-1)$-ary tree of depth $t-r$, $T_{k\to l}^{(t-r)}$. The message passed through $k\to l$ after $t-r$ iterations is distributed as $m^{(t-r)}$. 
This implies that, for consistency reasons, the probability distribution $P(m^{(t)})$ satisfies the following distributional equation
\begin{widetext}
    \begin{equation}
        P\left(m^{(t)}\right)= \int \prod_{j=1}^{d-1} \dd{m_j} P\left(m_j^{(t-1)}\right)\delta\qty[m^{(t)}-g^{\text{BP}}\qty(
        \{m_j^{(t-1)}\})]\,,
    \end{equation}
\end{widetext}
where $m_j^{(t-1)}$ are independent copies of $m^{(t-1)}$and $m^{(t)}=g^{\text{BP}}\qty(\{m_j^{(t-1)}\})$ is shorthand notation for the BP equation (\ref{eq: prior BP update equation}), with $j=1,\dots,d-1$.

Assuming that this equation has a fixed point and that $m^{(t)}$ converges to said fixed point, we obtain the cavity equation by substituting $P(m^{(t)})$ with $P(m)$:
\begin{equation}\label{eq: planted cavity equation}
    P(m)= \int \prod_{j=1}^{d-1} \dd{m_j} P(m_j) \delta\qty[m-g^{\text{BP}}\qty(\qty{m_j})] \,.
\end{equation}
Since the prior is a homogeneous Ising model defined on a random regular graph, there is no disorder in the prior and it is natural to look for a translational invariant solution to the BP equation.
The probability distribution $P(m)$ and the planted cavity equations simplify:
\begin{equation}
    P(m)=\delta[m-m^*] \qq{where} m^*=g^{\text{BP}}(m^*)\,.
\end{equation}

\subsection{Cavity equations for the posterior probability}

\subsubsection{Sampling the disordered variables}\label{app:sampling_disordered_variables_RS}
Applying the replica-symmetric cavity method directly to our posterior distribution (\ref{eq: posterior}) is not as straightforward since the BP equation (\ref{eq: posterior BP update equation}) depends on the couplings variables $\ud J$. 
In particular, these random variable are not independent, but are instead sampled from the probability distribution $P(\ud J)=\sum_{\ud s}P_{\beta}(\ud J|\ud s)P_{\kappa}(\ud s)$.
They do however become independent when conditioning on the planted configuration $\ud s$, since the likelihood (\ref{eq: likelihood}) factorizes.

Let us now consider once again $T_{i_0 \to j_0}^{(t)}$, in particular the couplings $\ud J_{T_{i_0 \to j_0}^{(t)}}=\{J_{ij}:\{i,j\}\in E(T_{i_0 \to j_0}^{(t)})\}$ and the spins $\ud s_{T_{i_0 \to j_0}^{(t)}}=\{s_{i}:i\in T_{i_0 \to j_0}^{(t)}\}$. Exploiting the tree structure, we can write 
\begin{align}
    P&\left(\ud J_{T_{i_0 \to j_0}^{(t)}}|\ud s_{T_{i_0 \to j_0}^{(t)}}\right)= \notag \\
    &\prod_{j=1}^{d-1} P\left(J_{j}|s_{i_0},s_j\right) P\left(\ud J_{T_{j \to i_0}^{(t-1)}}|\ud s_{T_{j \to i_0}^{(t-1)}}\right)\,,
\end{align}
where we denoted with $j=1,\dots , d-1$ the $d-1$ neighbors of $i_0$ and by $T_{j \to i_0}^{(t-1)}$ the directed sub-tree rooted in $j\to i_0$, which is a full $(d-1)$-ary tree of depth $t-1$.

As discussed in Appendix \ref{app:sampling_planted_config} we can use the BP estimates to sample the planted configuration $\ud s_{T_{i_0 \to j_0}^{(t)}}$:
\begin{align}
    P&\left(\ud s_{T_{i_0 \to j_0}^{(t)}}\right)=\int \dd{m}\int \dd{\ud m_{T_{i_0 \to j_0}^{(t)}}}\notag\\
    &P\left(\ud s_{T_{i_0 \to j_0}^{(t)}}| m, \ud m_{T_{i_0 \to j_0}^{(t)}}\right)P\left(m, \ud m_{T_{i_0 \to j_0}^{(t)}}\right)\,.
\end{align}
Here, $m$ denotes the limiting random variable to which the message $m^{(t)}$, defined in Appendix~\ref{app:RS_cavity_eq_prior}, converges in distribution as $t\to\infty$. 
The notation $\underline m_{T_{i_0\to j_0}^{(t)}}$ denotes the collection of analogous random variables associated with all edges of $T_{i_0\to j_0}^{(t)}$ oriented towards the root edge $i_0\to j_0$.
Using the fact that $T_{i_0 \to j_0}^{(t)}$ is a tree we can write
\begin{widetext}
    \begin{align}
    P\left(\ud s_{T_{i_0 \to j_0}^{(t)}}| m, \ud m_{T_{i_0 \to j_0}^{(t)}}\right)&=m(s_{i_0})\prod_{j=1}^{d-1} P(s_{j}| s_{i_0},m_{j})P\left(\ud s_{T_{j \to i_0}^{(t-1)}\setminus j}| s_j, m_{j},\ud m_{T_{j \to i_0}^{(t-1)}}\right)\,,\\
    P\left(m, \ud m_{T_{i_0 \to j_0}^{(t)}}\right)&=\delta[m- g^{\text{BP}}(\{m_{j}\})]\prod_{j=1}^{d-1}P\left(m_{j}, \ud m_{T_{j \to i_0}^{(t-1)}}\right)\,,\label{eq:sampling_prior_messages_RS}
\end{align}
\end{widetext}

\subsubsection{Cavity equations}\label{app:RS_cavity_eq_posterior}
We start by considering the message $\nu_{i_0\to j_0}^{(t)}$ going from node $i_0$ to $j_0$ at iteration $t$ of the BP algorithm and by assuming that the initial messages $\ud\nu^{(0)}$ are i.i.d. random variables. The message $\nu_{i_0\to j_0}^{(t)}$ will depend on the couplings
\begin{equation}
    \ud J_{B_{i_0\to j_0}^{(t)}}= \qty{J_{kl}: \{k,l\}\in E(B_{i_0\to j_0}^{(t)})}\,,
\end{equation}
and the initial messages 
\begin{equation}
    \ud \nu_{\partial B_{i_0\to j_0}^{(t)}}^{(0)}= \qty{\nu^{(0)}_{k\to p(k)}: k\in \partial B_{i_0\to j_0}^{(t)}}\,,
\end{equation}
where $B_{i_0\to j_0}^{(t)}$ is the directed neighborhood of $i_0\to j_0$ of depth $t$.
Similarly to the cavity equation derived for the prior in Appendix~(\ref{app:RS_cavity_eq_prior}), the neighborhood $B_{i_0\to j_0}^{(t)}$ converges in distribution to $T_{i_0 \to j_0}^{(t)}$, a full $(d-1)$-ary tree of depth $t$. It follows that, for $N\to \infty$, $\nu_{i_0 \to j_0}^{(t)}\xrightarrow{d}\nu^{(t)}$ where $\nu^{(t)}$ is the message going through the root of the tree $T_{i_0 \to j_0}^{(t)}$ at iteration $t$ of the BP algorithm. 

The message $\nu^{(t)}$ is an $N$-independent random variable that will depend on $\ud J_{T_{i_0 \to j_0}^{(t)}}$ and on $\ud\nu_{\partial T_{i_0 \to j_0}^{(t)}}^{(0)}=\qty{\nu_{k\to p(k)}^{(0)}: k \text{ leaf of } T_{i_0 \to j_0}^{(t)}} $, in particular: 
\begin{equation}\label{eq:nu_equation}
    \nu^{(t)}=f\left(\ud J_{T_{i_0 \to j_0}^{(t)}},\ud\nu_{\partial T_{i_0 \to j_0}^{(t)}}^{(0)}\right)\,,
\end{equation}
where $f$ is some function that can be computed by recursively applying (\ref{eq: posterior BP update equation}) starting from the leaves of the tree $T_{i_0 \to j_0}^{(t)}$.
Therefore, the probability distribution $P(\nu^{(t)})$ must satisfy 
\begin{widetext}
    \begin{align}
        P(\nu&^{(t)})=\sum_{\ud J_{T_{i_0 \to j_0}^{(t)}}} \int \dd{\ud \nu_{\partial T_{i_0 \to j_0}^{(t)}}^{(0)}} P\left(\ud J_{T_{i_0 \to j_0}^{(t)}},\ud \nu_{\partial T_{i_0 \to j_0}^{(t)}}^{(0)}\right)\delta \qty[\nu^{(t)}-f\left(\ud J_{T_{i_0 \to j_0}^{(t)}},\ud \nu_{\partial T_{i_0 \to j_0}^{(t)}}^{(0)}\right)]\notag\\
        &= \sum_{\ud s_{T_{i_0 \to j_0}^{(t)}}}\sum_{\ud J_{T_{i_0 \to j_0}^{(t)}}} \int \dd{\ud \nu_{\partial T_{i_0 \to j_0}^{(t)}}^{(0)}} P\left(\ud J_{T_{i_0 \to j_0}^{(t)}}|\ud s_{T_{i_0 \to j_0}^{(t)}}\right)P\left(\ud s_{T_{i_0 \to j_0}^{(t)}}\right)P\left(\ud \nu_{\partial T_{i_0 \to j_0}^{(t)}}^{(0)}\right)\delta \qty[\nu^{(t)}-f\left(\ud J_{T_{i_0 \to j_0}^{(t)}},\ud \nu_{\partial T_{i_0 \to j_0}^{(t)}}^{(0)}\right)]\notag \\
        &= \sum_{\ud s_{T_{i_0 \to j_0}^{(t)}}}\sum_{\ud J_{T_{i_0 \to j_0}^{(t)}}} \int \dd{\ud \nu_{\partial T_{i_0 \to j_0}^{(t)}}^{(0)}} \int \dd{ m}\int \dd{\ud m_{T_{i_0 \to j_0}^{(t)}}}P\left(\ud J_{T_{i_0 \to j_0}^{(t)}}|\ud s_{T_{i_0 \to j_0}^{(t)}}\right)\notag\\
        &\quad \times P\left(\ud s_{T_{i_0 \to j_0}^{(t)}\setminus i_0}| s_{i_0}, \ud m_{T_{i_0 \to j_0}^{(t)}}\right)m(s_{i_0})P\left(m, \ud m_{T_{i_0 \to j_0}^{(t)}}\right)P\left(\ud \nu_{\partial T_{i_0 \to j_0}^{(t)}}^{(0)}\right)\delta \qty[\nu^{(t)}-f\left(\ud J_{T_{i_0 \to j_0}^{(t)}},\ud \nu_{\partial T_{i_0 \to j_0}^{(t)}}^{(0)}\right)]\,.
    \end{align}
\end{widetext}
Writing the probability $P(\nu^{(t)})$ as the marginal of $P(\nu^{(t)},s_{i_0}, m)$, we can get rid of $m$ from both sides of the equation:
\begin{widetext}
    \begin{align}
        P(\nu&^{(t)}|s_{i_0}, m)P( m)= \sum_{\ud s_{T_{i_0 \to j_0}^{(t)}\setminus i_0}} \sum_{\ud J_{T_{i_0 \to j_0}^{(t)}}} \int\dd{\ud \nu_{\partial T_{i_0 \to j_0}^{(t)}}^{(0)}} \int\dd{\ud m_{T_{i_0 \to j_0}^{(t)}}} P\left(\ud J_{T_{i_0 \to j_0}^{(t)}}|\ud s_{T_{i_0 \to j_0}^{(t)}}\right)\notag\\
        &P\left(\ud s_{T_{i_0 \to j_0}^{(t)}\setminus i_0}| s_{i_0}, \ud m_{T_{i_0 \to j_0}^{(t)}}\right) P\left(m, \ud m_{T_{i_0 \to j_0}^{(t)}}\right) P\left(\ud \nu_{\partial T_{i_0 \to j_0}^{(t)}}^{(0)}\right) \delta\qty[\nu^{(t)}-f\left(\ud J_{T_{i_0 \to j_0}^{(t)}},\ud \nu_{\partial T_{i_0 \to j_0}^{(t)}}^{(0)}\right)]\,.\label{eq:recursion_element_RS_cavity_eq}
    \end{align}
\end{widetext}
Recognizing the terms introduced in Appendix \ref{app:sampling_disordered_variables_RS} the equation becomes 
\begin{widetext}
    \begin{align}
        &P\left(\nu^{(t)}|s_{i_0}, m\right)P(m)= \sum_{\ud s_{T_{i_0 \to j_0}^{(t)}\setminus i_0}}\sum_{\ud J_{T_{i_0 \to j_0}^{(t)}}} \int \dd{\ud \nu_{\partial T_{i_0 \to j_0}^{(t)}}^{(0)}}\int \prod_{j=1}^{d-1}\dd{m^{(t-1)}_j}\dd{\ud m_{T_{j \to i_0}^{(t-1)}}} \delta\left[m- g^{\text{BP}}(\{m_{i}\})\right] \notag \\
        &\delta \qty[\nu^{(t)}-f(\ud J_{T_{i_0 \to j_0}^{(t)}},\ud \nu_{\partial T_{i_0 \to j_0}^{(t)}}^{(0)})] P\left(\ud \nu_{\partial T_{i_0 \to j_0}^{(t)}}^{(0)}\right)\prod_{j=1}^{d-1} P\left(J_{j}|s_{i_0},s_j\right)P\left(\ud J_{T_{j \to i_0}^{(t-1)}}|\ud s_{T_{j \to i_0}^{(t-1)}}\right) P\left(s_{j}| s_{i_0},m_{j}\right)\notag \\
        &P\left(\ud s_{T_{j \to i_0}^{(t-1)}\setminus j}| s_j, m_{j},\ud m_{T_{j \to i_0}^{(t-1)}}\right) P\left(m_{j}, \ud m_{T_{j \to i_0}^{(t-1)}}\right)\,.
    \end{align}
\end{widetext}
Finally, we exploit the fact that $T_{i_0 \to j_0}^{(t)}$ is a tree
\begin{equation}
    P\left(\ud \nu_{\partial T_{i_0 \to j_0}^{(t)}}^{(0)}\right)=\prod_{j=1}^{d-1} P\left(\ud \nu_{\partial T_{j \to i_0}^{(t-1)}}^{(0)}\right)\,,
\end{equation} 
and we write the constraint enforcing the function (\ref{eq:nu_equation}) in terms of the BP update rule (\ref{eq: posterior BP update equation}) 
\begin{widetext}
    \begin{equation}
        \delta\left[\nu^{(t)}-f\left(\ud J_{T_{i_0 \to j_0}^{(t)}},\ud \nu_{\partial T_{i_0 \to j_0}^{(t)}}^{(0)}\right)\right]= \int\bigg(\prod_{j=1}^{d-1}\dd{\nu^{(t-1)}_j}\delta[\nu_{j}^{(t-1)}-f(\ud J_{T_{j \to i_0}^{(t-1)}},\ud\nu_{\partial T_{j \to i_0}^{(t-1)}}^{(0)})]\bigg) \delta[\nu^{(t)}-f^{\text{BP}}(\{\nu_{j}^{(t-1)},J_{j}\})]\,,
    \end{equation}
\end{widetext}
where $\nu_j^{(t-1)}$ are independent copies of $\nu^{(t-1)}$ and $\nu^{(t)}=f^{\text{BP}}\left(\{\nu_{j}^{(t-1)},J_{j}\}\right)$ is shorthand notation for the BP equation (\ref{eq: posterior BP update equation}), with $j=1,\dots,d-1$. 
The equation becomes
\begin{widetext}
    \begin{align}
        P&\left(\nu^{(t)}|s_{i_0}, m\right)P(m)=\int\prod_{j=1}^{d-1}\left(\dd{\nu^{(t-1)}_j}\dd{m_{j}}\right) \delta[m- g^{\text{BP}}(\{m_{j}\})]\sum_{\{ s_{j}, J_{j}\}} \prod_{j=1}^{d-1} P(J_{j}|s_{i_0},s_j) P(s_{j}| s_{i_0},m_{j})\notag \\
        & \times \delta[\nu^{(t)}-f^{\text{BP}}(\{\nu_{j}^{(t-1)},J_{j}\})]\sum_{\ud s_{T_{j \to i_0}^{(t-1)}\setminus j}}\sum_{\ud J_{T_{j \to i_0}^{(t-1)}}}\int \dd{\ud \nu_{\partial T_{j \to i_0}^{(t-1)}}^{(0)}}\int \dd{\ud m_{T_{j \to i_0}^{(t-1)}}}\delta\left[\nu_{j}^{(t-1)}-f\left(\ud J_{T_{j \to i_0}^{(t-1)}},\ud\nu_{\partial T_{j \to i_0}^{(t-1)}}^{(0)}\right)\right] \notag \\
        & \times P\left(\ud \nu_{\partial T_{j \to i_0}^{(t-1)}}^{(0)}\right)P\left(\ud J_{T_{j \to i_0}^{(t-1)}}|\ud s_{T_{j \to i_0}^{(t-1)}}\right) P\left(m_{j}, \ud m_{T_{j \to i_0}^{(t-1)}}\right)
        P\left(\ud s_{T_{j \to i_0}^{(t-1)}\setminus j}| s_j, m_{j},\ud m_{T_{j \to i_0}^{(t-1)}}\right) 
    \end{align}
\end{widetext}
By noticing the presence of a term analogous to (\ref{eq:recursion_element_RS_cavity_eq}) but with $T_{j \to i_0}^{(t-1)}$ instead of $T_{i_0 \to j_0}^{(t)}$, we can explicit the recursive nature of the equation
\begin{widetext}
    \begin{align}
        P(\nu^{(t)}|s_{i_0}, m)P( m)&=\int\prod_{j=1}^{d-1}\dd{m_{j}}P( m_{j}) \delta[m- g^{\text{BP}}(\{m_{j}\})]\sum_{\{ s_{j}, J_{j}\}} \prod_{j=1}^{d-1} P(J_{j}|s_{i_0},s_j) P(s_{j}| s_{i_0},m_{j}) \notag \\
        & \times\int\prod_{j=1}^{d-1}\dd{\nu^{(t-1)}_j}P(\nu^{(t-1)}_{j}|s_{j}, m_{j})\delta[\nu^{(t)}-f^{\text{BP}}(\{\nu_{j}^{(t-1)},J_{j}\})]
    \end{align}
\end{widetext}
The cavity equation is obtained by substituting $P(\nu^{(t)}|s_{i_0}, m)$ with the fixed point $P(\nu|s_{i_0}, m)$:
\begin{widetext}
    \begin{align}
        P(\nu|s, m)P( m)&=\int\prod_{j=1}^{d-1}\dd{m_{j}}P( m_{j})  \delta[m- g^{\text{BP}}(\{m_{j}\})] \sum_{\{ s_{j}, J_{j}\}} \prod_{j=1}^{d-1} P(J_{j}|s,s_j) P(s_{j}| s,m_{j}) \notag \\
        & \times\int\prod_{j=1}^{d-1}\dd{\nu_j}P(\nu_{j}|s_{j}, m_{j}) \delta[\nu-f^{\text{BP}}(\{\nu_{j},J_{j}\})]\label{eq:RS_cavity_eq_posterior}
    \end{align}
\end{widetext}
where we dropped the $i_0$ subscript.
For random regular graphs, and using the fact that the prior is a homogeneous Ising model, the probability $P(m)$ simplifies to a delta distribution and the cavity equation can be simplified to 
\begin{widetext}
    \begin{align} \label{eq: posterior RS cavity equation}
        &P(\nu|s, m^*)=\sum_{\{ s_{j}, J_{j}\}} \prod_{j=1}^{d-1} P(J_{j}|s,s_j) P(s_{j}| s,m^*)\int\dd{\nu_j}P(\nu_{j}|s_{j}, m^*) \delta[\nu-f^{\text{BP}}(\{\nu_{j},J_{j}\})]
    \end{align}
\end{widetext}
where $m^*=g^{\text{BP}}(m^*)$ ($g^{\text{BP}}$ being a shorthand notation for the BP equation (\ref{eq: prior BP update equation})).
When $\kappa\in [\kappa_{\text{rsb}},\kappa_c]$ the prior is in a paramagnetic phase and therefore $m^*(s)=1/2$. When this is the case, we have that $P(s_{j}| s,m^*)=\frac{e^{\kappa s s_j}}{2\cosh{\kappa}}$ and the dependence of $\nu$ on $m^*$ can be dropped:
\begin{widetext}
    \begin{align}\label{eq:RS_cavity_eq_paramagnetic_prior}
        &P(\nu|s)=\sum_{\{ s_{j}, J_{j}\}} \prod_{j=1}^{d-1} P(J_{j}|s,s_j) P(s_{j}| s)\int\dd{\nu_j}P(\nu_{j}|s_{j}) \delta[\nu-f^{\text{BP}}(\{\nu_{j},J_{j}\})]
    \end{align}
\end{widetext}
The distributional cavity equations (\ref{eq: posterior RS cavity equation}) and (\ref{eq:RS_cavity_eq_paramagnetic_prior}) can be solved numerically using the population dynamics algorithm.  The distributions $P(\nu|s,m^*)$ and $P(\nu|s)$ are approximated with a population of $N$ tuples $(\nu_i^{(+)},\nu_i^{(-)})$ and the equations are solved iteratively by interpreting them as update equations.

Let us now consider $\mu_{i_0}^{(t)}$, the estimated marginal of the node $i_0$ after $t$ iterations of the BP algorithm. By a similar argument as before, for $N\to \infty$ it will converge in distribution towards a random variable $\mu^{(t)}$, which is the BP estimate of the marginal of node $i_0$ on a full $d$ regular tree $T_{i_0}^{(t+1)}$ of depth $t+1$. All the nodes of the tree have degree $d$, except for the leaves. 
By the same logic as that used to obtain (\ref{eq:RS_cavity_eq_posterior}) we obtain
\begin{widetext}
    \begin{align}
        &P(\mu,s, b)=\int\prod_{j=1}^{d}\dd{m_{j}}P( m_{j})  \delta[b- \tilde g^{\text{BP}}(\{m_{j}\})] b(s)\sum_{\{ s_{j}, J_{j}\}} \prod_{j=1}^{d} P(J_{j}|s,s_j) P(s_{j}| s,m_{j}) \notag \\
        & \int\prod_{j=1}^{d}\dd{\nu_j}P(\nu_{j}|s_{j}, m_{j}) \delta[\mu-\tilde f^{\text{BP}}(\{\nu_{j},J_{j}\})]\,,\label{eq:RS_cavity_eq_marginal_posterior}
    \end{align}
\end{widetext}
where $b=\tilde g^{\text{BP}}(\{m_{j}\})$ and $\mu=\tilde f^{\text{BP}}(\{\nu_{j},J_{j}\})$ are shorthand notation for the BP marginal estimates (\ref{eq: prior BP marginal}) and (\ref{eq: posterior BP marginal equation}), respectively, with $j=1,\dots, d$. The previous simplifications of (\ref{eq:RS_cavity_eq_posterior}) also apply to this equation.

\subsubsection{Observables}\label{app: RS observables}

The RS prediction of the free entropy is obtained by averaging the Bethe free entropy (\ref{eq:bethe_free_entropy_posterior}) over the probability distribution of the messages: 
\begin{equation}
    \label{eq:free_entropy_RS} 
    \Phi^{\text{RS}}=\mathbb{E}^{\text{RS}}[\ln z_i] -\dfrac{d}{2}\mathbb{E}^{\text{RS}}[\ln z_{ij}]\,.
\end{equation}
It can be easily shown that the two elements can be written as
\begin{widetext}
    \begin{align}
        \mathbb{E}^{\text{RS}}[\ln z_i]&=\sum_s \int\dd b \int(\prod_{j=1}^{d}\dd{m_{j}}P( m_{j}))  \delta[b- \tilde g^{\text{BP}}(\{m_{j}\})] b(s)\sum_{\{ s_{j}, J_{j}\}_{j=1}^{d}} (\prod_{j=1}^{d} P(J_{j}|s,s_j) P(s_{j}| s,m_{j})) \notag \\
        & \times\int(\prod_{j=1}^{d}\dd{\nu_j}P(\nu_{j}|s_{j}, m_{j})) \ln z_i(\{\nu_{j},J_{j}\})\,,\\
        \mathbb{E}^{\text{RS}}[\ln z_{ij}]&=\int\dd{m_i}\int\dd{m_j} P(m_i) P(m_j) \sum_{s_i,s_j, J_{ij}} P(J_{ij}|s_i,s_j) P(s_j|s_i, m_j) P(s_i|m_i,m_j)\notag \\
        & \times\int\dd{\nu_i}\int\dd{\nu_j}P(\nu_i, |s_i, m_i)P( \nu_j|s_j,m_j)\ln z_{ij}(\nu_i,\nu_j,J_{ij})\,,
    \end{align}
\end{widetext}
where we explicited the dependence of the normalization factors $z_i$ and $z_{ij}$ on the messages and the couplings.
The other observables of interest are the overlaps
\begin{align}
    \mathbb{E}_{\ud J,\ud \sigma, \ud \sigma'}[O(\ud \sigma,\ud \sigma')] &= \dfrac{1}{N}\sum_{i=1}^{N}  \mathbb{E}_{\ud J}[ (\sum_{\sigma_i}P_{\beta,\kappa,i}( \sigma_i|\ud J)\sigma_i)^2]\,,\label{eq: overlap estimator}\\
    \mathbb{E}_{\ud J,\ud \sigma, \ud s}[O(\ud \sigma,\ud s)] &=  \dfrac{1}{N}\sum_{i=1}^{N}  \mathbb{E}_{\ud J, \ud s}[ \sum_{\sigma_i}P_{\beta,\kappa,i}( \sigma_i|\ud J)\sigma_i s_i]\,,\\
    \mathbb{E}_{\ud J,\ud \sigma, \ud s}[O(\ud{\hat{\sigma}},\ud s)] &=  \dfrac{1}{N}\sum_{i=1}^{N}  \mathbb{E}_{\ud J}[ \hat\sigma_i s_i]\,.
\end{align}
The RS estimates of the overlaps are all calculated in a similar manner, so we can focus on only one of the three. We consider the overlap (\ref{eq: overlap estimator}) between two configurations sampled from the posterior (\ref{eq: posterior}). It can easily be shown that the overlap depends on the marginals
\begin{align}\label{eq:overlap_config_step}
    \mathbb{E}_{\ud J,\ud \sigma, \ud \sigma'}[O(\ud \sigma,\ud \sigma')] &= \dfrac{1}{N}\sum_{i=1}^{N}  \sum_{\ud J} P(\ud J) \qty(\sum_{\sigma_i}P_{\beta,\kappa,i}( \sigma_i)\sigma_i)^2.
\end{align}
Let us consider the estimated marginal $\mu_i^{(t)}$ after $t$ iterations, which we know converges in distribution to the random variable $\mu_i^{(t)}$. As stated in the previous section, the marginal $\mu_i^{(t)}$ is a function of the couplings $\ud J$ and the initial messages $\ud \nu^{(0)}_{\partial T_{i}(t+1)}$. Therefore we can write 
\begin{align}
    P(\mu_i^{(t)})=&\sum_{\ud J}\int\dd{\ud \nu^{(0)}_{\partial T_{i}(t+1)}} P(\ud \nu^{(0)}_{\partial T_{i}(t+1)}) P(\ud J)\notag \\
    &\times\delta[\mu_i^{(t)}-\tilde f(\ud \nu^{(0)}_{\partial T_{i}(t+1)},\ud J_{T_{i}(t+1)} ))]\,,
\end{align}
where $\tilde f$ is some function that can be computed from (\ref{eq: posterior BP marginal equation}) and then recursively applying (\ref{eq: posterior BP update equation}) starting from the leaves of the tree $T_{i}(t+1)$.

Introducing the initial messages $\ud \nu^{(0)}_{\partial T_{i}(t+1)}$, the delta distribution in (\ref{eq:overlap_config_step}) and taking the limit $t\to \infty$, we find
\begin{align}
    \mathbb{E}^{\text{RS}}[O(\ud \sigma,\ud \sigma')] &= \int\dd{\mu}P(\mu) \qty(\sum_{\sigma}\mu( \sigma)\sigma)^2\,.
\end{align}
Finally, we write the probability distribution $P(\mu)$ as a marginal of (\ref{eq:RS_cavity_eq_marginal_posterior}) and find
\begin{align}
    \mathbb{E}^{\text{RS}}[O(\ud \sigma,\ud \sigma')] &= \sum_s\iint\dd b\dd{\mu}P(\mu, s, b) \qty(\sum_{\sigma}\mu( \sigma)\sigma)^2\,.
\end{align}

Similarly
\begin{align}
    \mathbb{E}^{\text{RS}}[O(\ud \sigma,\ud s)] &=\iint\dd{b}\dd{\mu}\sum_{s} P(\mu, s, b)s\sum_\sigma\mu(\sigma)\sigma \\
    \mathbb{E}^{\text{RS}}[O(\ud s,\ud{\hat \sigma})] &= \iint\dd{b}\dd{\mu}\sum_{s} P(\mu, s, b)s \hat \sigma\,.
\end{align}

\subsection{Stability analysis}
\label{app:stability_analysis}
In the region $\kappa\in [\kappa_{\text{rsb}},\kappa_c]$ the cavity equation (\ref{eq:RS_cavity_eq_paramagnetic_prior}) admits the trivial paramagnetic fixed point
\begin{equation}
    P(\nu| s)=\delta\qty[\nu-\bar \nu] \qq{with} \bar \nu(\sigma)=\dfrac{1}{2}\,.
\end{equation}

This fixed point is the one found by the BP algorithm when the all the messages in the initial condition are set to $1/2$. We wish to study its stability with respect to a small perturbation: if this is the case, any initial condition close to it will lead to the trivial fixed point and therefore this will be the only solution the BP algorithm will be able to find.

We can study the stability by considering a distribution $P(\nu| s)$, solution of the distributional cavity equation, close to the paramagnetic one: 
\begin{equation}
    \nu_{k\to i}(\sigma_k)= \dfrac{1}{2}+\varepsilon_{k\to i}(\sigma_k)\,,
\end{equation}
where $\varepsilon_{k\to i}(\sigma_k)$ is a small perturbation such that $\sum_{\sigma_k}\varepsilon_{k\to i}(\sigma_k)=0$.
Introducing this expression in the BP equation (\ref{eq: posterior BP update equation}) and keeping only linear terms in $\varepsilon$ we find
\begin{equation}
    \nu_{i\to j}(\sigma_i)= \dfrac{1}{2}+\sum_{k\in \partial i\setminus j} \sum_{\sigma_k}\dfrac{e^{(\kappa+\beta J_{ik})\sigma_i \sigma_k}}{2\cosh{(\kappa+\beta J_{ik})} } \varepsilon_{k\to i}(\sigma_k)\,.
\end{equation}

The first moment of the distance between the paramagnetic solution and the perturbed solution:
\begin{equation}
    M_s(\sigma)=\int \dd{\nu} P(\nu|s)\qty(\nu(\sigma)- \dfrac{1}{2})\,,\label{eq: mean distance}
\end{equation}
quantifies the distance of the RS solution $P(\nu|s)$ from the trivial fixed point. 

Introducing the RS equation (\ref{eq:RS_cavity_eq_paramagnetic_prior}) and utilizing the normalization condition $\sum_{\sigma}M_s (\sigma)=0$ we find
\begin{equation}
    M_s= (d-1)\sum_{J} \dfrac{\tanh{(\kappa+\beta J)}}{4\cosh{\kappa}\cosh{\beta}} \sum_{s'} e^{(\kappa+\beta J)s's} M_{s'}\,,
\end{equation}
where we defined $M_s(+)\equiv M_s$ for simplicity of notation

We can write this in the form of a linear system
\begin{equation}
    \ud M^{(t+1)}= A\ud M^{(t)}\,,
\end{equation}
where 
\begin{equation}
    \ud M = \pmqty{M_+\\ M_-},\quad A=\mqty(a &  b \\ b & a)\,,
\end{equation}
and
\begin{align}
    a&=(d-1) \sum\limits_{J} \dfrac{\tanh{(\kappa+\beta J)}}{4\cosh{\kappa}\cosh{\beta}} e^{\kappa+\beta J},\\
    b&= (d-1) \sum\limits_{J} \dfrac{\tanh{(\kappa+\beta J)}}{4\cosh{\kappa}\cosh{\beta}} e^{-(\kappa+\beta J)}.
\end{align}
The paramagnetic fixed point will be stable if and only if both the eigenvalues are smaller than one. It is easily shown that this corresponds to 
\begin{align}
    \kappa&<\operatorname{atanh}\qty(\dfrac{1}{d-1})=\kappa_c\,, \\
    1&>(d-1)\dfrac{\sum_{J}\tanh{(\kappa+\beta J)}\sinh{(\kappa+\beta J)}}{2\cosh{\kappa} \cosh{\beta}}\,.
\end{align}
The first condition enforces the prior to be in the paramagnetic phase, while the second condition delimitates the impossible phase for the posterior (see Figure~\ref{fig:phase_diagram}, region 1.).

\section{One-step Replica Symmetry Breaking cavity method}
\label{app:1RSB_formalism}
For $\kappa<\kappa_{\rm rsb}$, the prior~\eqref{eq: prior} enters a static RSB phase, where the Gibbs measure decomposes into a convex combination of Bethe states whose number grows sub-exponentially with $N$.
In this regime, the RS cavity method is not sufficient anymore and must be replaced by the 1RSB cavity formalism, which can be summarized as follows: introducing a Boltzmann distribution over Bethe states, writing this measure in the form of a graphical model, and using BP to study the resulting model.
We define the 1RSB messages as
\begin{equation}
    Q_{i\to j}(\nu_{i\to j})=\mathbb{P}_{c\sim  p(c)}\qty{\nu^c_{i\to j}=\nu_{i\to j}}\,,
\end{equation}
i.e. the probability to observe a BP message $\nu^c_{i\to j}=\nu_{i\to j}$ with $c$ chosen randomly from the set of Bethe measures $\cal C$. 
The update equation for such messages is 
\begin{widetext}
    \begin{equation}\label{eq: RSB BP equation posterior}
        Q_{i\to j}(\nu_{i \to j})=\dfrac{1}{{Z}_{i\to j}}\int\qty(\prod_{k\in \partial i\setminus j}\dd{\nu_{k\to i}} Q_{k\to i}(\nu_{k \to i}))\delta\qty[\nu_{i \to j}- f^{\text{BP}}(\ud\nu_{\partial i\setminus j},\ud J_{\partial i\setminus j})]\qty(z_{i\to j}(\ud\nu_{\partial i\setminus j},\ud J_{\partial i\setminus j}))^x\,,
    \end{equation}
\end{widetext}
where $\ud\nu_{\partial i\setminus j}=\{\nu_{k\to i}:k\in \partial i\setminus j\}$ and $\ud J_{\partial i\setminus j}=\{J_{ik}:k\in \partial i\setminus j\}$, $Z_{i\to j}$ is the normalization factor and $x$ is the Parisi parameter.
The corresponding equation for the prior is obtained by setting $\beta =0$. We denote the 1RSB messages of the prior as $R_{i\to j}$, the normalization factor ${Z}^0_{i\to j}$ and the Parisi parameter $x_0$:
\begin{widetext}
    \begin{equation}\label{eq: RSB BP equation prior}
        R_{i\to j}(m_{i \to j})=\dfrac{1}{{Z}^0_{i\to j}}\int\qty(\prod_{k\in \partial i\setminus j}\dd{m_{k\to i}} R_{k\to i}(m_{k \to i}))\delta\qty[m_{i \to j}- g^{\text{BP}}(\ud m_{\partial i\setminus j})]\qty(z^0_{i\to j}(\ud m_{\partial i\setminus j}))^{x_0}\,.
    \end{equation}
\end{widetext}

It follows that the marginal estimates $\tilde Q_{i}$ and $\tilde R_i$, of the posterior and the prior respectively, are
\begin{widetext}
    \begin{align}
        \tilde Q_{i}(\mu_i)&=\dfrac{1}{{Z}_{i}}\int\qty(\prod_{k\in \partial i}\dd{\nu_{k\to i}} Q_{k\to i}(\nu_{k \to i}))\delta\qty[\mu_{i }- \tilde f^{\text{BP}}(\ud\nu_{\partial i},\ud J_{\partial i})]\qty(z_{i}(\ud\nu_{\partial i},\ud J_{\partial i}))^x \,, \label{eq: RSB BP marginal posterior}\\
        \tilde R_{i}(b_{i})&=\dfrac{1}{{Z}^0_{i}}\int\qty(\prod_{k\in \partial i}\dd{m_{k\to i}} R_{k\to i}(m_{k \to i}))\delta\qty[b_{i}- \tilde g^{\text{BP}}(\ud m_{\partial i})]\qty(z^0_{i}(\ud m_{\partial i}))^{x_0}\,.\label{eq: RSB BP marginal prior}
    \end{align}
\end{widetext}
where $\ud\nu_{\partial i}=\{\nu_{k\to i}:k\in \partial i\}$, $\ud m_{\partial i}=\{m_{k\to i}:k\in \partial i\}$, $\ud J_{\partial i}=\{J_{ik}:k\in \partial i\}$, and $Z_{i},\, Z_{i}^0$ are the normalization factors.

\subsection{Cavity equations for the prior probability}
\subsubsection{Cavity equations}
Obtaining the 1-RSB cavity equation for the prior probability is analogous to what done in Appendix \ref{app:RS_cavity_eq_prior}. The message $R_{i_0 \to j_0}^{(t)}$ passing through the directed edge $i_0\to j_0$ at iteration $t$ will depend on the initial messages $\ud R^{(0)}_{\partial B_{i_0 j_0}^{(t)}}$ and will converge in distribution to a random variable $R^{(t)}$. When $t\to \infty$, $R^{(t)}$ converges in distribution to the random variable $R$, whose probability distribution satisfies 
\begin{equation}\label{eq: planted RSB cavity equation}
    P(R)=\int \qty(\prod_{j=1}^{d-1} \dd{R_j} P(R_j))\delta\qty[R-g^{\text{RSB}}(\qty{R_{j}})]\,,
\end{equation}
where $\{R_j\}_{j\in\{1,\dots,d-1\}}$ are independent copies of $R$ and $R=g^{\text{RSB}}(\qty{R_{j}})$ is shorthand notation for the BP update rule (\ref{eq: RSB BP equation prior}).

This is a self-consistent equation of a probability distributions over another probability distribution, which makes it very cumbersome to solve, even numerically. In particular, to implement the population dynamics algorithm one would have to represent $P(R)$ with a sample of distributions $\{R_i\}_{i=1}^{N}$, each of which has to be encoded by a population of messages $\{m_i,j\}_{j=1}^{N}$.

However, in the case of regular graphs and with a homogeneous coupling constant $\kappa$, the probability distribution $P(R)$ takes the form $P(R)=\delta\qty[R-R^*]$
where
\begin{widetext}
    \begin{equation}\label{eq:prior_RSB_FP_eq}
        R^*(m)=\int\prod_{j=1}^{d-1}\dd{m_{j}} R^*(m_{j})\delta\qty[m- g^{\text{BP}}(\qty{m_{j}})]\dfrac{\qty(z^0(\qty{m_j}))^{x_0}} {Z^0}\,,
    \end{equation}
\end{widetext}
which is a fixed point equation of a probability distribution ($Z^0$ being the normalization factor in\eqref{eq: RSB BP equation prior}).

\subsubsection{Accounting for the re-weighting factors}
The re-weighting factor $z^0/Z^0$ in the fixed point equation (\ref{eq:prior_RSB_FP_eq}) implies that, at each population-dynamics iteration, the messages $\{m^{(t)}_j\}$ cannot simply be sampled independently: they must instead be drawn from the corresponding re-weighted joint distribution.

A standard way do it is through a re-weighting procedure~\cite{Mezard_2009}, by constructing an intermediate population by sampling the messages $\{m^{(t)}_j\}$ independently, storing for each element the value of $z^0$ and then constructing the final population by sampling from the intermediate population using the $z^0$ as weights. However, for $\kappa<-\kappa_c$ the dynamics becomes trapped by the attractor
\begin{equation}
    R^{(t)}(m)=\delta[m-m_+]\,, \qq{} R^{(t+1)}(m)=\delta[m-m_-]\,,
\end{equation}
with 
\begin{equation}
    m_{\pm}(s)=\dfrac{1\pm s u(\kappa)}{2}\,,
\end{equation}
for some $u(\kappa)>0$.

To circumvent this problem, we use an accept-reject procedure instead. We sample the messages $\{m^{(t)}_j\}$ independently, sample a random variable $z_{\text{sampled}}^0\sim\operatorname{Unif}([0,z_{\text{max}}^0])$, with $z_{\text{max}}^0=2\cosh[(d-1)\kappa ]$, and accept the proposed message $m^{(t+1)}$ only if $(z_{\text{sampled}}^0)^{x_0}<(z^0)^{x_0}$.

\subsubsection{Estimating the Parisi parameter $x_0$}
\label{subsubsec:estimating_x0}
\begin{figure}[t]
	\centering
	\includegraphics[width=\linewidth]{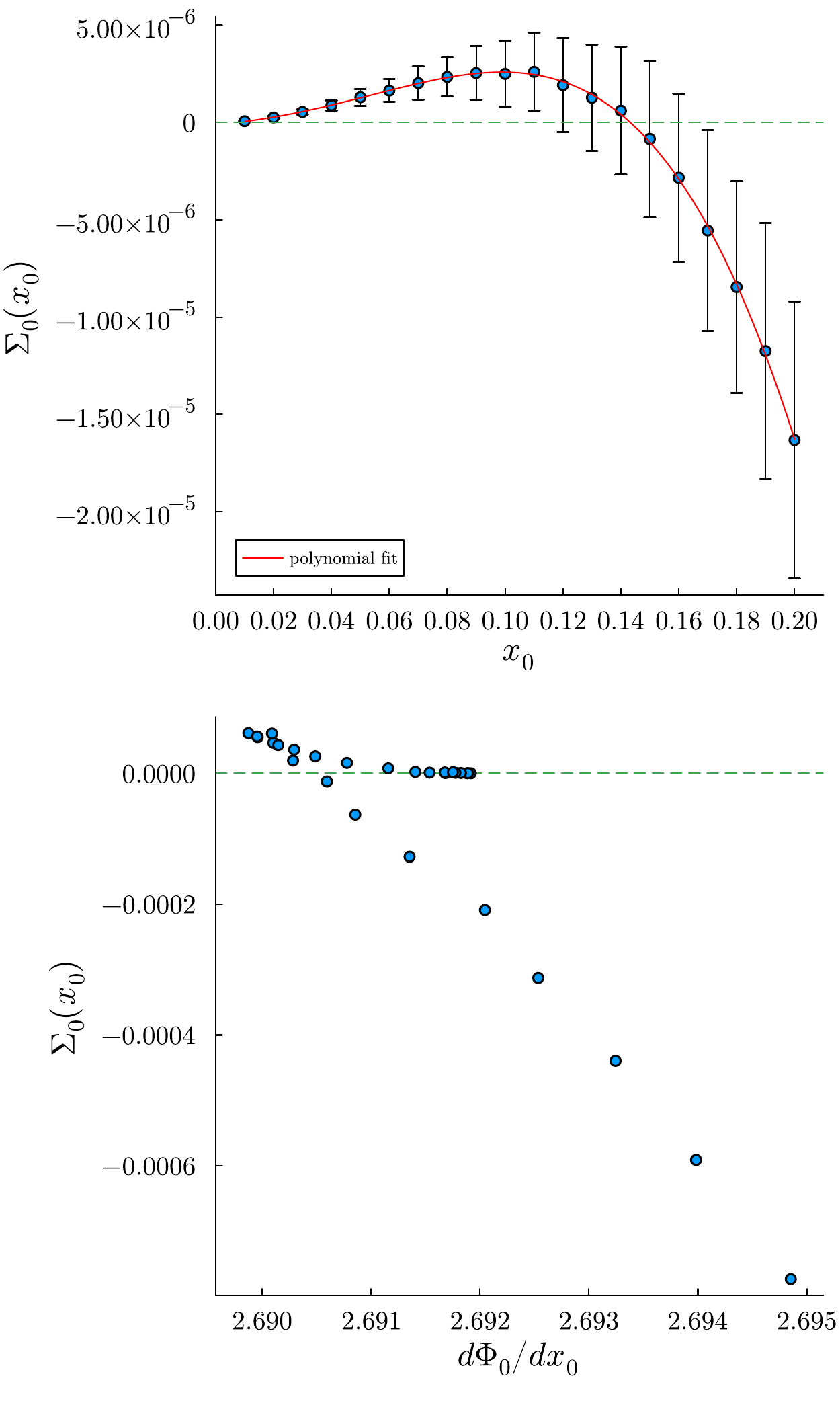}
	\caption{
	Top: Plot of the complexity of the prior $\Sigma_0(x_0)$ with respect to the Parisi parameter $x_0$, for $\kappa=-1.2$. 
	Bottom: Corresponding parametric plot of $\Sigma_0(x_0)$ versus the replicated free-entropy derivative ${\rm d}\Phi_0^{\rm rsb}/{\rm d}x_0$ (varying $x_0\in[0,1]$). 
	These results were obtained by running a population dynamics algorithm for the 1RSB cavity equation~\eqref{eq:prior_RSB_FP_eq} at $\kappa=-1.2$ (top panel) and $\kappa=-2.1$ (bottom panel), with population $N=10^6$.}
	\label{fig:x0_fit_kappa-1.2}
\end{figure}
In the 1RSB formalism, the physical free-entropy of the prior, denoted $\Phi_0$, is obtained via the Laplace method:
\begin{equation}
	\label{eq:physical_potential}
	\Phi_0=\sup_{\phi: \Sigma_0(\phi)\geq0}\left[\Sigma_0(\phi)+\phi\right]\,,
\end{equation}
where $\Sigma_0(\phi)$ is the complexity, i.e. $1/N$ times the log of the number of clusters having free-entropy $\phi$.
The condition $\Sigma_0(\phi)\geq0$ ensures that the clusters exist in the thermodynamic limit.
If the value $\phi^*$ achieving the supremum is such that $\Sigma_0(\phi^*)>0$, then there is an exponential number of clusters contributing to the total free entropy density, corresponding to a dynamical RSB phase. 
If $\Sigma_0(\phi^*)=0$ instead, there is only a sub-exponential number of dominant clusters, corresponding to a static RSB phase.

In the 1RSB cavity formalism, one introduces the potential $\Phi_0^{\rm rsb}$ called the replicated entropy:
\begin{align}
	\label{eq:1RSB_potential}
	\begin{aligned}
	\Phi_0^{\rm RSB}(x_0)&=\sup_\phi\left[\Sigma_0(\phi)+x_0\phi\right]\\
	&=\Sigma_0(\phi^*(x_0))+x_0\phi^*(x_0)\,,
	\end{aligned}
\end{align}
where $\phi^*(x_0)$ is the value achieving the supremum.
Evaluating $\Phi_0^{\rm rsb}(x_0)$ allows to compute $\Phi_0$ (while taking care of the constraint $\Sigma(\phi)\geq0$), and in turn to compute all physical observables.
The potential $\Phi_0^{\rm rsb}$ and the complexity are Legendre transform of each other, and one can therefore invert this relation to get:
\begin{align}
	\begin{aligned}
	\Sigma_0(\phi^*(x_0))&=\Phi_0^{\rm rsb}(x_0)-x_0\phi^*(x_0)\\
	&=\Phi_0^{\rm rsb}(x_0)-x_0\frac{{\rm d}\Phi_0^{\rm rsb}(x_0)}{{\rm d}x_0}\,,
	\end{aligned}
\end{align}
The potential $\Phi_0^{\rm rsb}(x_0)$ and its derivative can be computed with the 1RSB formalism, leading to an estimation of the complexity function $\Sigma_0(x_0)=\Sigma_0(\phi^*(x_0))$, see Figure~\ref{fig:x0_fit_kappa-1.2}, top panel, for such function plotted at $\kappa=-1.2$.

In a dynamical RSB phase, the supremum $\phi^*$ of~(\ref{eq:physical_potential}) is achieved for $\Sigma_0(\phi^*)>0$, therefore $\phi^*=\phi^*(x_0=1)$ and $\Phi_0=\Phi_0^{\rm rsb}(x_0=1)$. 
It can be moreover checked that $\Phi_0^{\rm rsb}(x_0=1)$ coincides with the RS prediction of the free-entropy: therefore in a dynamic RSB transition all observables coincide with the RS prediction. The RSB phenomenon is however detected by the slow down of the dynamics (hence the name dynamic RSB phase).

In a static RSB phase instead (as it is the case for the prior~\eqref{eq: prior}, for $\kappa<\kappa_{\rm rsb}$), the supremum $\phi^*$ is such that $\Sigma_0(\phi^*)=0$, and $\phi^*=\phi^*(\hat{x}_0)$ for a smaller value $0<\hat{x}_0<1$.
In Figure~\ref{fig:x0_fit_kappa-1.2} (top panel), one can indeed see that $\Sigma_0(x_0=1)<0$: therefore $x_0=1$ is not the correct value, and $\hat{x}_0$ is selected as the value at which $\Sigma_0$ vanishes.

Note that the complexity $\Sigma_0$ is not a monotonous function of $x_0$, and tends to $0$ as $x_0\to0$.
This observation can be understood by looking at the parametric plot $\Sigma_0(x_0)$ versus $\phi^*(x_0)={\rm d}\Phi_0^{\rm rsb}(x_0)/{\rm d}x_0$ obtained by varying $x_0\in[0,1]$ (shown in Figure~\ref{fig:x0_fit_kappa-1.2}, bottom panel). 
Starting from $x_0=1$ and decreasing towards zero, one first moves on the lower branch, where the complexity $\Sigma_0$ increases until it reaches its maximum at the cusp. 
Then, decreasing further $x_0$, one then moves on the upper branch, where the complexity decreases towards zero.

We solve the 1RSB equation for the prior~(\ref{eq:prior_RSB_FP_eq}) with a population size $N=10^6$  (or $10^7$ for the $\kappa$ values close to the critical threshold) performing $maxiter = 1000$ iterations. For a fixed value of $\kappa$ we repeated the process for different values of $x_0$ in $[0,1]$, performed a polynomial fit and looked for the value $\hat{x}_0$ at which $\Sigma_0(\hat{x}_0) =0$. Figure \ref{fig:x0_fit_kappa-1.2} (top panel) shows such a fit for $\kappa=-1.2$. 

The relevant values of the complexity are extremely small (of order $10^{-6}$ for $\kappa=-1.2$). This scale decreases further as $\kappa$ approaches $\kappa_{\rm rsb}$, making the numerical estimation of $\hat{x}_0$ increasingly delicate near the transition.

\subsection{Cavity equations for the posterior probability}
\subsubsection{Sampling the disordered variables}
Once again, the couplings $\ud J$ are not independent. To account for this we must condition on the planted configuration, which must be sampled by in turn introducing the planted messages.
Differently to the RS case, the planted BP messages $\{m_{j\to i_0}\}_{j\in\{1,\dots,d-1\}}$ incoming into node $i_0$ cannot be considered independent and therefore one must be careful in the way they are sampled.
We can sample the planted BP messages recursively, starting from the root of the truncated tree $T_{i_0\to j_0}^{(t)}$:
$$
P(m_{i_0\to j_0}|\underline{R_{i_0\to j_0}})=R_{i_0\to j_0}(m_{i_0\to j_0})\,,
$$
Then we write the conditional probabilities for the $d-1$ neighboring messages:
\begin{widetext}
    \begin{equation}
        P(\{m_{j\to i}\}|m_{i_0\to j_0},\{R_{j\to i}\})= \dfrac{\delta[m_{i_0\to j_0}-g^{\text{BP}}(\{m_{j\to i}\})] (z^0(\{m_{j\to i}\}))^{x_0}\prod_{j=1}^{d-1}R_{j\to i}(m_{j\to i})}{Z^0 (\{R_{j\to i}\})g^{\text{RSB}}(\{R_{j\to i}\})}\,.
    \end{equation}
\end{widetext}
To sample a set of planted BP messages on the tree $T_{i_0\to j_0}^{(t)}$, we therefore start at the root and sample $m\sim R(m)$, and then sample the BP messages of the children of $i_0$ from this conditional probability, and go down to the leaves in this way.
The set of 1RSB messages $\{R_{i\to j}\}$ are sampled under the RS approximation. 
This effectively means replacing (\ref{eq:sampling_prior_messages_RS}) with
\begin{widetext}
    \begin{align}
        P\left(m, \ud m_{T_{i_0 \to j_0}^{(t)}}\right)&=\int\dd{R}\int \dd{R_{T_{i_0 \to j_0}^{(t)}}} P\left(m, \ud m_{T_{i_0 \to j_0}^{(t)}}|R, \ud R_{T_{i_0 \to j_0}^{(t)}}\right)P\left(R, \ud R_{T_{i_0 \to j_0}^{(t)}}\right)\,,\\
    P\left(m, \ud m_{T_{i_0 \to j_0}^{(t)}}|R, \ud R_{T_{i_0 \to j_0}^{(t)}}\right)&=R(m)P(\{m_{j}\}|m,\{R_{j}\})\prod_{j=1}^{d-1}P\left(m_{T_{j \to i_0}^{(t-1)}}|m_{j}, \ud R_{T_{j \to i_0}^{(t-1)}}\right)\,,\\
    P\left(R, \ud R_{T_{i_0 \to j_0}^{(t)}}\right)&=\delta[R- g^{\text{RSB}}(\{R_{j}\})]\prod_{j=1}^{d-1}P\left(R_{j}, \ud R_{T_{j \to i_0}^{(t-1)}}\right)\,.
    \end{align}
\end{widetext}
Once the set of planted BP messages are drawn, we sample the planted configuration $\underline{s}$ on the tree in the same way as we did for the RS cavity equations, using the tree-like structure, conditioned on the planted BP messages.

\subsubsection{Cavity equations}\label{app:sampling_disordered_variables_RSB}
Obtaining the 1-RSB cavity equation for the prior probability is analogous to what was done in Appendix \ref{app:RS_cavity_eq_posterior}.

The message $Q_{i_0\to j_0}^{(t)}$ going from node $i_0$ to $j_0$ at iteration $t$ of the BP algorithm depends on the couplings $\ud J_{B_{i_0\to j_0}^{(t)}}$ and the initial messages $\ud Q_{\partial B_{i_0\to j_0}^{(t)}}^{(0)}$ and converges in distribution to the message going through the root of the tree $T_{i_0 \to j_0}^{(t)}$ at iteration $t$ of the BP algorithm $Q_{i_0\to j_0}^{(t)}$. 

Following the steps outlined in Appendix \ref{app:RS_cavity_eq_posterior} and sampling the disordered variables following Appendix \ref{app:sampling_disordered_variables_RSB} we obtain 
\begin{widetext}
    \begin{align}\label{eq: 1RSB distributional cavity equation posterior}
    P(Q|s,m,R)P(R)&= \int\prod_{j=1}^{d-1}\dd{R_{j}}P(R_{j})\delta[R-g^{\text{RSB}}(\{R_{j}\})]\int\prod_{j=1}^{d-1}\dd{m_{j}} P(\{m_{j}\} | m,\{R_{j}\})\times \notag 
    \\
    &\times \sum_{\{J_j, {s_j}\}} \prod_{j=1}^{d-1} P(J_i | {s_j},s) P(s_j| s, m_{j}) \int\prod_{j=1}^{d-1}\dd{Q_{j}}P(Q_{j}|s_j, m_{j}, R_{j})\delta[Q-f^{\text{RSB}}(\{{Q_{j}}, J_j\})]\,,
    \end{align}
\end{widetext}
where $Q=f^{\text{RSB}}(\qty{{Q_{j}}, J_j})$ is shorthand notation for (\ref{eq: RSB BP equation posterior}).
For random regular graphs and with homogeneous prior coupling constant $\kappa$, the equation simplifies to 
\begin{widetext}
    \begin{align} \label{eq: 1RSB distributional cavity equation posterior rnd reg graphs}
    P(Q|s,m,R^*)&= \int\prod_{j=1}^{d-1}\dd{m_{j}} P(\{m_{j}\} | m,R^*)\sum_{\{J_j, {s_j}\}} [\prod_{j=1}^{d-1} P(J_j | {s_j},s) P(s_j| s, m_{j}) \notag \\
    &\times \int\prod_{j=1}^{d-1}\dd{Q_{j}}P(Q_{j}|s_j, m_{j}, R^*)\delta[Q-f^{\text{RSB}}(\{{Q_{j}}, J_j\})]\,,
    \end{align}
\end{widetext}
where $R^*=g^{\text{RSB}}(R^*)$ is a shorthand notation for~\eqref{eq:prior_RSB_FP_eq}.

Let us now consider $\tilde Q_{i_0}^{(t)}$, the estimated marginal of the node $i_0$ after $t$ iterations of the BP algorithm. By a similar argument as before, for $N\to \infty$ it will converge in distribution towards a random variable $\tilde Q^{(t)}$, which is the BP estimate of the marginal of node $i_0$ on a full $d$ regular tree $T_{i_0}^{(t+1)}$ of depth $t+1$. All the nodes of the tree have degree $d$, except for the leaves. 

By the same logic as that used to obtain (\ref{eq: 1RSB distributional cavity equation posterior rnd reg graphs}) we obtain
\begin{widetext}
    \begin{align}
    P(\tilde{Q},s,b,\tilde R) &= \int(\prod_{j=1}^{d}\dd{R_{j}}\delta[R_j-R^*])\delta[\tilde{R}-g^{\text{RSB}}(\{R_j\})]\tilde{R}(b)\int(\prod_{j=1}^{d}\dd{m_{j}}) P(\{m_{j}\} | b,R^*)\delta[b-\tilde{g}^{\text{BP}}(\qty{m_j})]b(s) \notag \\
    &\times \sum_{\qty{J_j, {s_j}}} \qty[\prod_{j=1}^{d} P\qty(J_j \mid {s_j},s) P\qty(s_j \mid s, m_{j}) ]\int\qty(\prod_{j=1}^{d}\dd{Q_{j}}P(Q_{j}|s_{j},m_{j},R_j)) \delta[\tilde Q-f^{\text{RSB}}(\{{Q_{j}}, J_j\})]\,,\label{eq: 1RSB marginal cavity equation posterior rnd reg graphs}
\end{align} 
\end{widetext}
where $\tilde Q=f^{\text{RSB}}(\{{Q_{j}}, J_j\})$ is shorthand notation for the BP marginal estimates (\ref{eq: RSB BP marginal posterior}), with $j=1,\dots, d$.

\subsubsection{Simplifications at $x=1$}
\label{app: x=1 simplifications}

Similarly to (\ref{eq: planted RSB cavity equation}), the 1RSB distributional cavity equation (\ref{eq: 1RSB distributional cavity equation posterior}) is very complicated to solve numerically and would need to introduce populations of populations. Furthermore, the probability distribution of $Q$ is conditioned on the messages $R$, which are themselves probability distributions.

The equation can be simplified greatly for $x=1$, and the dynamic and static RSB phase transition thresholds can be obtained through the RSB approach at this specific value of $x$.
Let us define the average message 
\begin{equation}
    \bar \nu[Q]=\int \dd{\nu} Q(\nu) \nu \,.
\end{equation}
One can easily check that the normalization of the 1RSB equation (\ref{eq: RSB BP equation posterior}) is $Z= z(\qty{\bar \nu[Q_j],J_j}_{j=1,\dots,d-1})$ where $z$ is the normalization of the BP equation (\ref{eq: posterior BP update equation}).

We define the probability distribution $P(\bar \nu|s,m,R) \equiv  \mathbb{P}\qty{\bar \nu[Q]=\bar \nu|s,m,R}$, which can be easily shown to obey the RS equation with an RSB prior:
\begin{widetext}
    \begin{align}\label{eq: RS distributional cavity equation with RSB prior}
    P(\bar \nu | s,m,R)P(R)&= \int(\prod_{j=1}^{d-1}\dd{R_{j}}P(R_{j}))\delta[R-g^{\text{RSB}}(\{R_{j}\})]\int(\prod_{j=1}^{d-1}\dd{m_{j}}) P(\{m_{j}\} | m,\{R_{j}\})\times \notag \\
    &\times\sum_{\{J_j, {s_j}\}} [\prod_{j=1}^{d-1} P(J_j |{s_j},s) P(s_j| s, m_{j}) ]\int(\prod_{j=1}^{d-1}\dd{\bar \nu_{i}}P(\bar \nu_{j}|s_j, m_{j}, R_{j}))\delta[\bar \nu-f^{\text{BP}}(\{{\bar \nu_{j}}, J_j\})]\,.
    \end{align}
And on random regular graphs this equation becomes
	\begin{align}\label{eq: RS distributional cavity equation with RSB prior_RR}
		P(\bar \nu | s,m,R^*)&= \int(\prod_{j=1}^{d-1}\dd{m_{j}}) P(\{m_{j}\} | m,\{R^*\})\times \notag \\
		&\times\sum_{\{J_j, {s_j}\}} [\prod_{j=1}^{d-1} P(J_j |{s_j},s) P(s_j| s, m_{j}) ]\int(\prod_{j=1}^{d-1}\dd{\bar \nu_{i}}P(\bar \nu_{j}|s_j, m_{j}, R^*))\delta[\bar \nu-f^{\text{BP}}(\{{\bar \nu_{j}}, J_j\})]\,.
	\end{align}
\end{widetext}
We can then define the average distribution of the message $\nu$ conditioned on its average value $\bar \nu$:
\begin{widetext}
    \begin{equation}\label{eq: average distribution conditioned on the average value of the message}
        \bar Q(\nu|\bar \nu,s,m,R) P(R)= \dfrac{1}{P(\bar \nu|s,m,R)} \int \dd{Q} P(Q|s,m,R)P(R) Q(\nu) \delta[\bar \nu-\bar \nu[Q]]\,,
    \end{equation}
\end{widetext}
for which we can write a closed form equation:
\begin{widetext}
    \begin{align}
    \bar Q(\nu|\bar \nu,s,m,R)&P(\bar \nu | s,m,R)P(R)= \int(\prod_{j=1}^{d-1}\dd{R_{j}}P(R_{j}))\delta[R-g^{\text{RSB}}(\{R_{j}\})]\int(\prod_{j=1}^{d-1}\dd{m_{j}}) \notag \\
    &\times P(\{m_{j}\} | m,\{R_{j}\})\sum_{\{J_j, {s_j}\}} [\prod_{j=1}^{d-1} P(J_j |{s_j},s) P(s_j| s, m_{j}) ]\int(\prod_{j=1}^{d-1}\dd{\bar \nu_{i}}P(\bar \nu_{j}|s_j, m_{j}, R_{j}))\delta[\bar \nu-f^{\text{BP}}(\{\bar \nu_{j}, J_j\})]\notag \\
    &\times \int(\prod_{j=1}^{d-1}\dd{ \nu_{j}}\bar Q( \nu_{j}|\bar \nu_j, s_j, m_{j}, R_{j}))\delta[\nu-f^{\text{BP}}(\{ \nu_{j}, J_j\})]\dfrac{z(\{\nu_j,J_j\})}{z(\{\bar \nu_j,J_j\})}\,,
\end{align}
\end{widetext}
This equation is much simpler than (\ref{eq: 1RSB distributional cavity equation posterior}) as $\bar Q$ is a distribution over $\nu$ instead of a distribution of distributions like $P(Q)$. 
The re-weighting factor ${z(\qty{\nu_j,J_j})}/{z(\qty{\bar \nu_j,J_j})}$ can be removed by defining
\begin{equation}
    \bar{Q}_\sigma(\nu|\bar \nu,s,m,R)=\dfrac{\nu(\sigma)}{\bar \nu (\sigma)}\bar Q(\nu|\bar \nu,s,m,R)\,,
\end{equation}
which can be easily shown to be normalized and thus can be interpreted as a probability distribution.
The simplified equation without the re-weighting factor is
\begin{widetext}
    \begin{align}
        \bar Q_{\sigma}(\nu|\bar \nu,s,m,R)&P(\bar \nu|s,m,R)P(R)= \int(\prod_{j=1}^{d-1}\dd{R_{j}}P(R_{j}))\delta[R-g^{\text{RSB}}(\{R_{j}\})]\int(\prod_{j=1}^{d-1}\dd{m_{j}}) P(\{m_{j}\} | m,\{R_{j}\}) \notag \\
        &\times\sum_{\{J_j, {s_j}\}} [\prod_{j=1}^{d-1} P(J_j | {s_j},s) P(s_j | s, m_{j}) ]\int(\prod_{j=1}^{d-1}\dd{\bar \nu_{i}}P(\bar \nu_{j}|s_{j},m_{j}, R_j))\delta[\bar \nu-f^{\text{BP}}(\{\bar \nu_{j}, J_j\})] \notag \\
        &\times\sum_{\{\sigma_j\}} [\prod_{j=1}^{d-1} P(\sigma_j | \sigma, J_{j}, {\bar \nu_{j}}) ] \int(\prod_{j=1}^{d-1}\dd{ \nu_{j}}\bar Q_{\sigma_j}(\nu_{j}|\bar \nu_{j},s_{j},m_{j}, R_{j}))\delta[ \nu-f^{\text{BP}}(\{ \nu_{j}, J_j\})]\,,
    \end{align}
\end{widetext}
where $P(\sigma_j | \sigma, J_{j}, {\bar \nu_{j}})=\dfrac{e^{(\kappa+\beta J_{j})\sigma \sigma_j}\bar\nu_j(\sigma_j)}{\sum_{\sigma_jd}e^{(\kappa+\beta J_{j})\sigma \sigma_j}\bar\nu_j(\sigma_j)}$.
Finally, we multiply both sides by $R(m)$ and use the fact that for random regular graphs and with constant prior coupling $\kappa$, the distribution $P(R)$ is a delta distribution to obtain 
\begin{widetext}
    \begin{align}\label{eq: simplified 1RSB distributional cavity equation}
        \bar Q_{\sigma}(\nu|\bar \nu,s,m)&P(\bar \nu|s,m)R^*(m)=\int(\prod_{i=1}^{d-1}\dd{m_{j}}R^*(m_i)) \delta[m-g^{\text{BP}}(\{m_{j}\})] \dfrac{ [z^0(\{m_{j}\})]^{x_0}}{Z^0}  \notag \\
        &\sum_{\qty{J_i, {s_i}}} \bigg[\prod_{j=1}^{d-1} P(J_i | {s_i},s)P(s_i | s, m_{j}) \bigg]\int(\prod_{j=1}^{d-1}\dd{\bar \nu_{j}}P(\bar \nu_{j}|s_{j},m_{j}))\delta[\bar \nu-f^{\text{BP}}(\{\bar \nu_{j}, J_i\})]\notag\\
        & \sum_{\{\sigma_i\}} [\prod_{j=1}^{d-1} P(\{\sigma_i | \sigma, J_{j}, \bar \nu_{j}\}) ] \int(\prod_{j=1}^{d-1}\dd{ \nu_{j}}\bar Q_{\sigma_i}(\nu_{j}|\bar \nu_{j},s_{j},m_{j})) \delta[ \nu-f^{\text{BP}}(\{ \nu_{j}, J_i\})]\,.
    \end{align}
\end{widetext}
This equation can be solved numerically using the population dynamics algorithm, with the population consisting of tuples of 7 elements: $(m_i,\{\bar\nu_i^{(s)}\},\{\nu_{i}^{(s,\sigma)}\})$.
We re-parametrized the messages as follows
\begin{align}
	\begin{aligned}
    m(s)=\dfrac{1+su}{2}\,, \qq{} \bar\nu(\sigma)=\dfrac{1+\sigma \bar h}{2}\,, \qq{} \nu(\sigma)=\dfrac{1+\sigma h}{2}\,,
	\end{aligned}
\end{align}
and used the following initial conditions
\begin{align}
    P_1(\bar h |s,u)&=\delta(\bar h -s)\,, \label{eq: first IC}\\
    P_2(\bar h |s,u)&=\delta(\bar h -1)\,,\\
    P_3(\bar h |s,u)&=\varepsilon\delta(\bar h -s)+(1-\varepsilon)\delta(h-u)\,, \\
    P_4(\bar h |s,u)&=\varepsilon\delta(\bar h -s)+(1-\varepsilon)\mathcal{U}([-1,1])\,,\\
    P_5(\bar h |s,u)&=(1-\varepsilon)\delta(\bar h)+\varepsilon\mathcal{U}([-1,1])\,, \label{eq: sRSB initial condition}
\end{align}
with $\varepsilon=0$ and $\varepsilon=0.001$.

\subsubsection{Observables}\label{app: RSB observables}
The complexity estimate at $x=1$ can be easily shown to be 
\begin{equation}
    \Sigma(x=1)=\Phi^{\text{RS}}-x(\phi^{\text{RSB}}_1 -\dfrac{d}{2}\phi^{\text{RSB}}_2)\,,
\end{equation}
with
\begin{widetext}
    \begin{align}        
    &\Phi^{\text{RS}}=\int\qty(\prod_{j=1}^{d}\dd{m_j}R^*(m_j))\dfrac{(z_i^0)^{x_0}}{\mathcal{Z}_i^0 }\sum_{s, \qty{s_j, J_j}}\qty[\prod_{j=1}^{d}P(J_j|s_j,s)P(s_j|s,m_j)]\int\qty(\prod_{j=1}^{d}\dd{\bar \nu_j}P(\bar \nu_j \mid s_j, m_j)) \notag \\
    &\times \tilde{g}^{BP}(\qty{m_j}, s)  \ln{z_i(\qty{\bar \nu_j, J_j})}-\dfrac{d}{2}\int\dd{m_j}\dd{m_j}R^*(m_i)R^*(m_j)\dfrac{(z_{ij}^0)^{x_0}}{\mathcal{Z}_{ij}^0 }\sum_{s, s_i,s_j,J_{ij}}P(J_{ij}|s_j,s_i)\notag \\
    &\times P(s_j|s_{i},m_j)\int\dd{\bar \nu_i}\dd{\bar \nu_j}P(\bar \nu_i \mid s_i, m_i)P(\bar \nu_j \mid s_j, m_j) P(s_i|m_i,m_j)  \ln{z_{ij}(\bar \nu_i,\bar \nu_j, J_{ij})}\,,\\  
    &\phi^{\text{RSB}}_1= \int\qty(\prod_{j=1}^{d}\dd{m_j}R^*(m_j))\dfrac{(z_i^0)^{x_0}}{\mathcal{Z}_i^0 }\sum_{s, \qty{s_j, J_j}}\qty[\prod_{j=1}^{d}P(J_j|s_j,s)P(s_j|s,m_j)]\int\qty(\prod_{j=1}^{d}\dd{\bar \nu_j}P(\bar \nu_j \mid s_j, m_j)) \notag \\
    &\times \sum_{\sigma, \qty{\sigma_j}}\qty[\prod_{j=1}^{d}P(\sigma_j|J_j,\bar \nu_j, \sigma)] \int\qty(\prod_{j=1}^{d}\dd{\nu_j}P(\nu_j \mid \bar \nu_j, s_j, m_j)) \tilde{g}^{BP}(\qty{m_j}, s) \tilde{f}^{BP}(\qty{\nu_j, J_j}, \sigma) \ln{z_i}\,,\\        
    &\phi^{\text{RSB}}_2= \int \dd{m_i } \dd{m_j} R^*(m_i)R^*(m_j)\dfrac{(z_{ij}^0)^{x_0}}{\mathcal{Z}_{ij}^0 }\sum_{s_i, s_j, J_{ij}}P(J_{ij}|s_i,s_j)P(s_j|s_i,m_j)P(s_i|m_i, m_j)\int\dd{\bar \nu_i}\dd{\bar \nu_j} \notag \\
    &\times P(\bar \nu_i \mid s_i, m_i)P(\bar \nu_j \mid s_j, m_j) \sum_{\sigma_i ,\sigma_j}P(\sigma_j|J_{ij},\bar \nu_j, \sigma_i) P(\sigma_i\mid J_{ij},\bar \nu_i, \bar \nu_j) \int\dd{\nu_i}\dd{\nu_j}P(\nu_i \mid \bar \nu_i, s_i, m_i)\notag \\
    &P(\nu_j \mid \bar \nu_j, s_j, m_j) \ln{z_{ij}}\,,
\end{align}
\end{widetext}
where the symbols with $\sim$ correspond to the quantities computed for the marginal probabilities (e.g. $\tilde{f}^{\rm BP}$ is the marginal BP update, $\tilde{Q}$ is the 1RSB marginal of the posterior).

The overlaps of interest are the inter-state and intra-state overlaps $q_0$ and $q_1$, as well as the overlaps of the planted configuration with a configuration sampled from the posterior and with the MMO estimator, $q_p$ and $q_e$ respectively. Following a similar procedure to that of Appendix \ref{app: RS observables} it can be shown that the RSB estimates are given by
\begin{widetext}
    \begin{align}
        q_0&= \int\dd{\tilde{Q}}\dd{b}\dd{\tilde{R}}\sum_s P(\tilde{Q},s,b,\tilde{R}) \qty(\int \dd{\mu}\tilde{Q}(\mu)\sum_\sigma  \mu(\sigma)\sigma)^2 \,, \\
        q_1&= \int\dd{\tilde{Q}}\dd{b}\dd{\tilde{R}}\sum_s P(\tilde{Q},s,b,\tilde{R}) \int \dd{\mu}\tilde{Q}(\mu)\qty(\sum_\sigma \mu(\sigma)\sigma)^2 \,,\\
        q_p&= \int\dd{\tilde{Q}}\dd{b}\dd{\tilde{R}}\sum_s P(\tilde{Q},s,b,\tilde{R})\int \dd{\mu}\tilde{Q}(\mu)\sum_\sigma  \mu(\sigma)\sigma s \,, \\
        q_e&= \int\dd{\tilde{Q}}\dd{b}\dd{\tilde{R}}\sum_s P(\tilde{Q},s,b,\tilde{R}) \int \dd{\mu}\tilde{Q}(\mu)\qty(\arg\max_{\sigma}\mu(\sigma)s) \,.
    \end{align}
\end{widetext}

One can introduce the expression (\ref{eq: 1RSB marginal cavity equation posterior rnd reg graphs}) of the probability distribution $P(\tilde{Q},s,b,\tilde{R})$ and can the repeat the steps outlined in Appendix \ref{app: x=1 simplifications} to obtain the overlaps from the conditional averaged messages $\bar Q_\sigma$: 
\begin{widetext}
    \begin{align}
    q_0 &= \int\qty(\prod_{j=1}^{d}\dd{m_{j}}R^*(m_j))\dfrac{ [z_i^0(\qty{m_{j}})]^{x_0}}{Z_i^0} \tilde{g}^{\text{BP}}(\qty{m_j},s)\sum_{s,\qty{J_j, {s_j}}} \qty[\prod_{j=1}^{d} P\qty(J_j \mid {s_j},s) P\qty(s_j \mid s, m_{j}) ]\notag \\
    & \int\qty(\prod_{j=1}^{d}\dd{\bar \nu_{j}}P(\bar \nu_{j}|s_{j},m_{j}))\qty(\sum_{\sigma}\tilde{f}^{\text{BP}}(\qty{{ \bar \nu_{j}}, J_j}, \sigma) \sigma)^2 \,, \\
    q_1&= \int\qty(\prod_{j=1}^{d}\dd{m_{j}}R^*(m_j)) \dfrac{ [z_i^0(\qty{m_{j}})]^{x_0}}{ Z_i^0}\tilde{g}^{\text{BP}}(\qty{m_j},s) \sum_{s,\qty{J_j, {s_j}}} \qty[\prod_{j=1}^{d} P\qty(J_j \mid {s_j},s) P\qty(s_j \mid s, m_{j}) ]\notag \\
    & \int\qty(\prod_{j=1}^{d}\dd{\bar \nu_{j}}P(\bar \nu_{j}|s_{j},m_{j})) \tilde{f}^{\text{BP}}(\qty{{ \bar \nu_{j}}, J_j}, \sigma) \sum_{\sigma,\qty{\sigma_j}} \qty[\prod_{j=1}^{d} P\qty(\qty{\sigma_j} \mid \sigma, J_{j}, {\bar \nu_{j}}) ]  \int\qty(\prod_{j=1}^{d}\dd{ \nu_{j}}\bar Q_{\sigma_j}(\nu_{j}|\bar \nu_{j},s_{j},m_{j}))
    \notag \\
    &\qty(\sum_{\sigma'}\tilde{f}^{\text{BP}}(\qty{{ \nu_{j}}, J_j}, \sigma') \sigma')^2 \,, \\
    q_p&= \int\qty(\prod_{j=1}^{d}\dd{m_{j}}R^*(m_j)) \dfrac{ [z_i^0(\qty{m_{j}})]^{x_0}}{ Z_i^0}\tilde{g}^{\text{BP}}(\qty{m_j},s)\sum_{s,\qty{J_j, {s_j}}} \qty[\prod_{j=1}^{d} P\qty(J_j \mid {s_j},s) P\qty(s_j \mid s, m_{j}) ]\notag \\
    & \int\qty(\prod_{j=1}^{d}\dd{\bar \nu_{j}}P(\bar \nu_{j}|s_{j},m_{j}))\qty(\sum_{\sigma}\tilde{f}^{\text{BP}}(\qty{{ \bar \nu_{j}}, J_j}, \sigma) \sigma s ) \,, \\
    q_e&= \int\qty(\prod_{j=1}^{d}\dd{m_{j}}R^*(m_j)) \dfrac{ [z_i^0(\qty{m_{j}})]^{x_0}}{Z_i^0}\tilde{g}^{\text{BP}}(\qty{m_j},s) \sum_{s,\qty{J_j, {s_j}}} \qty[\prod_{j=1}^{d} P\qty(J_j \mid {s_j},s) P\qty(s_j \mid s, m_{j}) ]\notag \\
    & \int\qty(\prod_{j=1}^{d}\dd{\bar \nu_{j}}P(\bar \nu_{j}|s_{j},m_{j}))\tilde{f}^{\text{BP}}(\qty{{ \bar \nu_{j}}, J_j}, \sigma)\sum_{\sigma,\qty{\sigma_j}} \qty[\prod_{j=1}^{d} P\qty(\qty{\sigma_j} \mid \sigma, J_{j}, {\bar \nu_{j}}) ]  \int\qty(\prod_{j=1}^{d}\dd{ \nu_{j}}\bar Q_{\sigma_j}(\nu_{j}|\bar \nu_{j},s_{j},m_{j}))
    \notag \\
    & \qty(\arg\max_{\sigma'}\tilde{f}^{\text{BP}}(\qty{{ \nu_{j}}, J_j}, \sigma')s)\,.
\end{align}
\end{widetext}

\subsection{Simplifications for Erd\H{o}s-R\'{e}nyi random graphs}
When the prior is not defined on a random regular graph, for instance on an Erd\H{o}s-R\'{e}nyi graph, or when its couplings are not homogeneous, the simplification $P(R)=\delta[R-R^*]$ no longer applies. 
One can nevertheless obtain tractable equations by setting $x_0=1$ and applying simplifications analogous to those used for the posterior in Appendix~\ref{app: x=1 simplifications}. 
This description would be correct if the prior exhibits only dynamical RSB, for which the thermodynamic solution remains at $x_0=1$.

\section{Supplementary plots}
\label{app:supplementary_plots}
In this appendix we report additional results obtained on the prior paramagnetic phase (Figure~\ref{fig:supp_plot_para}), ferromagnetic phase (Figure~\ref{fig:supp_plot_ferro}) and static RSB phase (Figures~\ref{fig:supp_plot_sRSB_1} and~\ref{fig:supp_plot_sRSB_2}). More plots can be found on the Github repository~\cite{github_ref}.
\begin{widetext}
	\begin{center}
		\begin{figure}[ht]
			\includegraphics[width=0.35\linewidth]{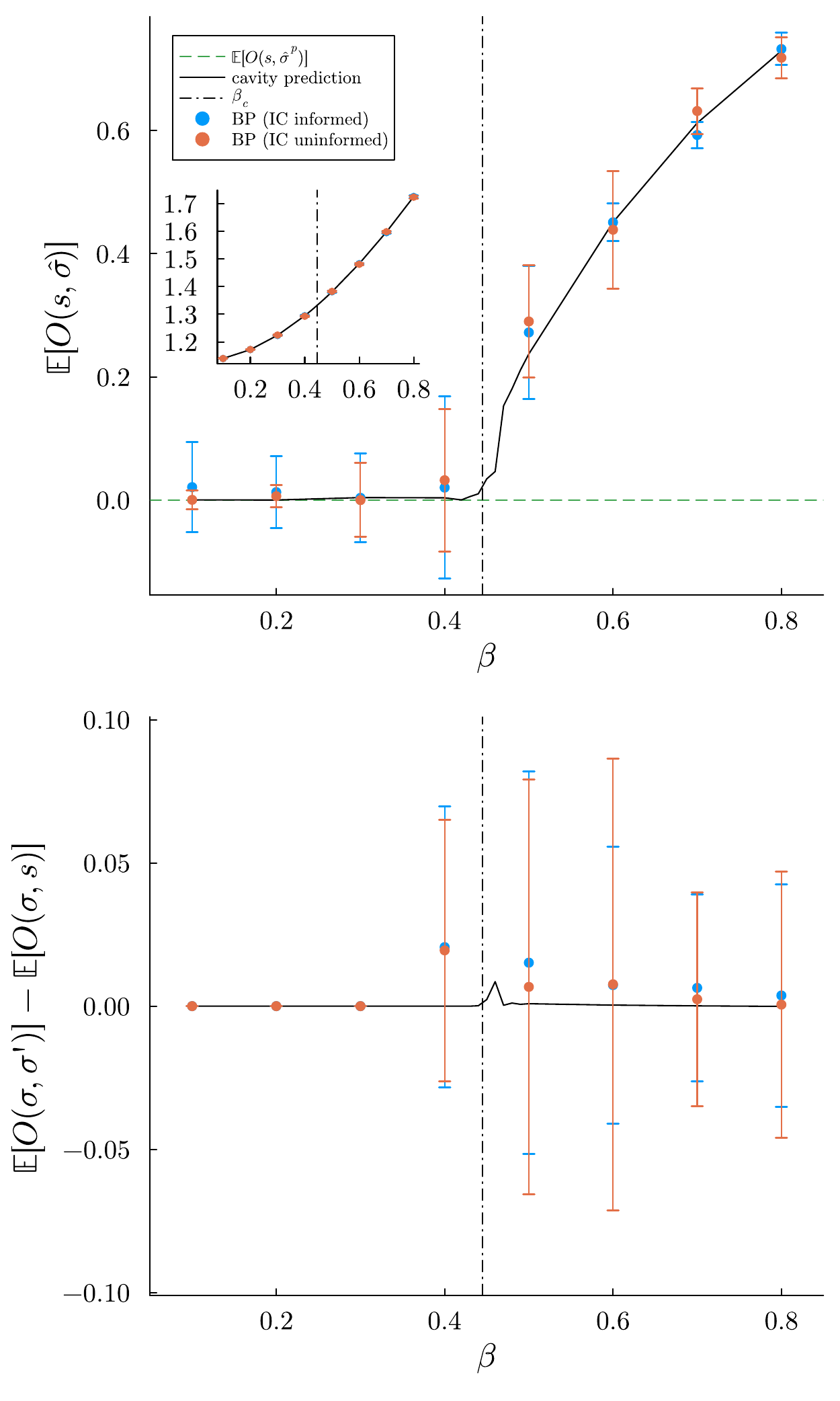}
			\includegraphics[width=0.35\linewidth]{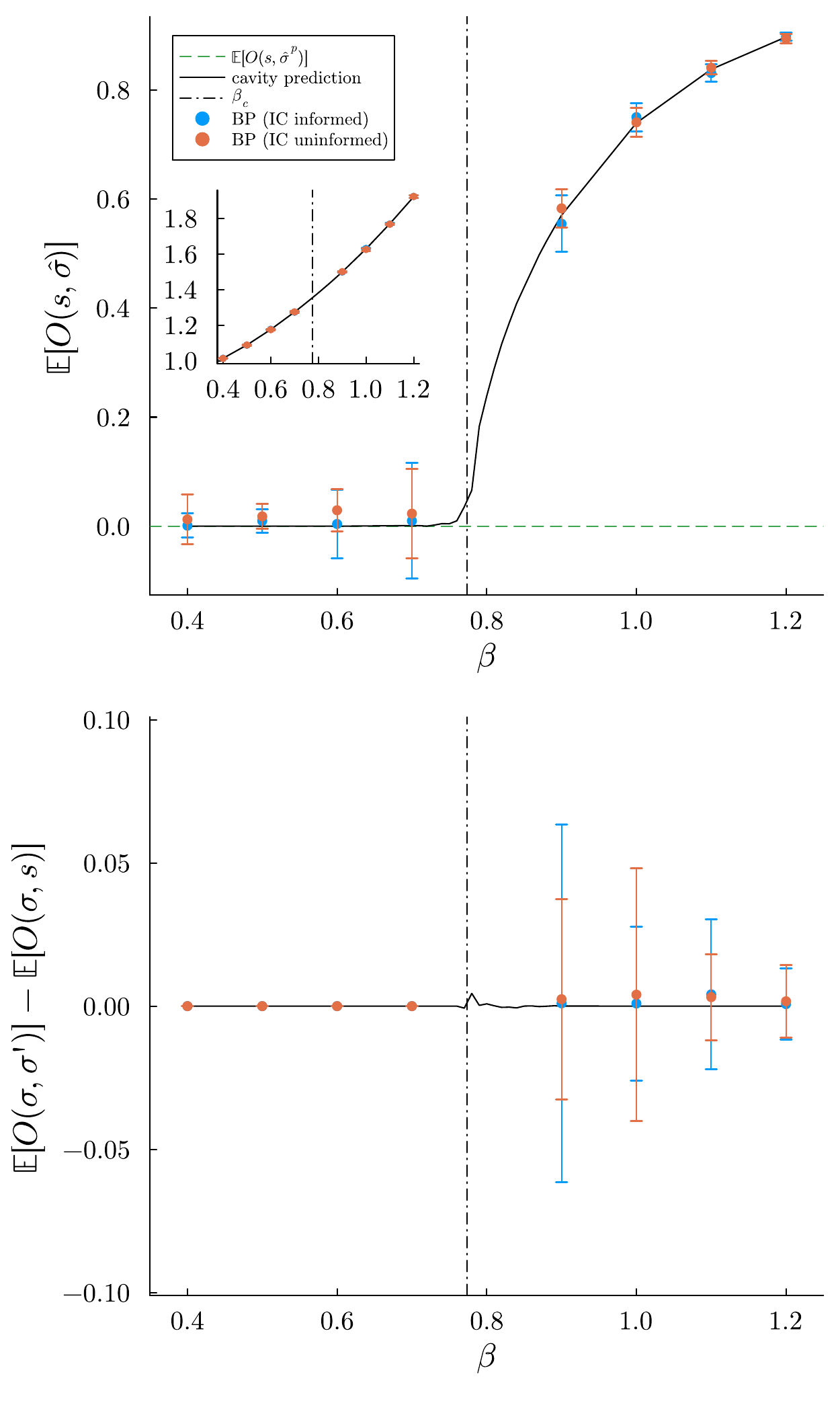}
			\caption{Results with paramagnetic prior. Left: $\kappa=-0.8$, Right: $\kappa=0.5$}
			\label{fig:supp_plot_para}
		\end{figure}
		\begin{figure}[h!]
			\includegraphics[width=0.35\linewidth]{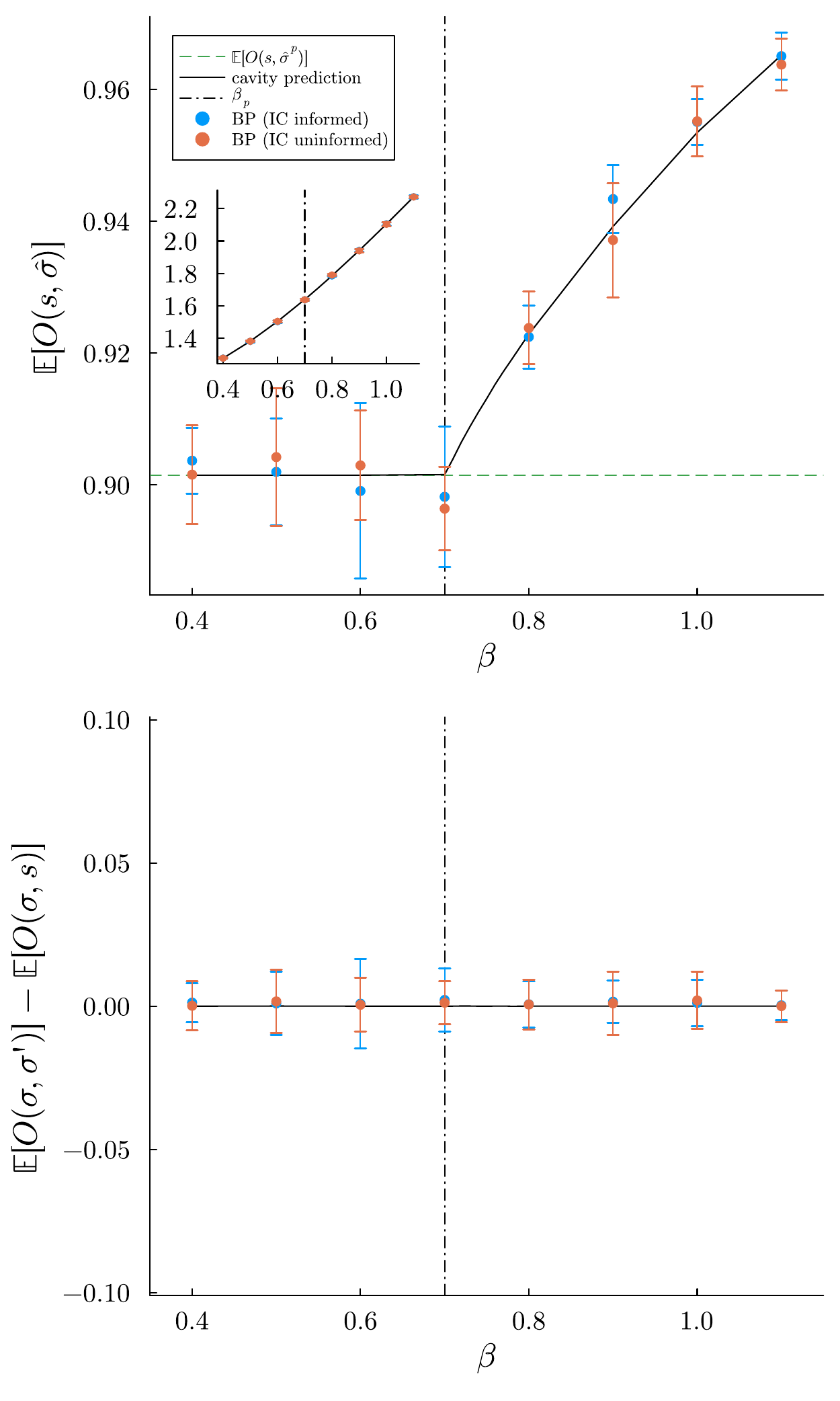}
			\includegraphics[width=0.35\linewidth]{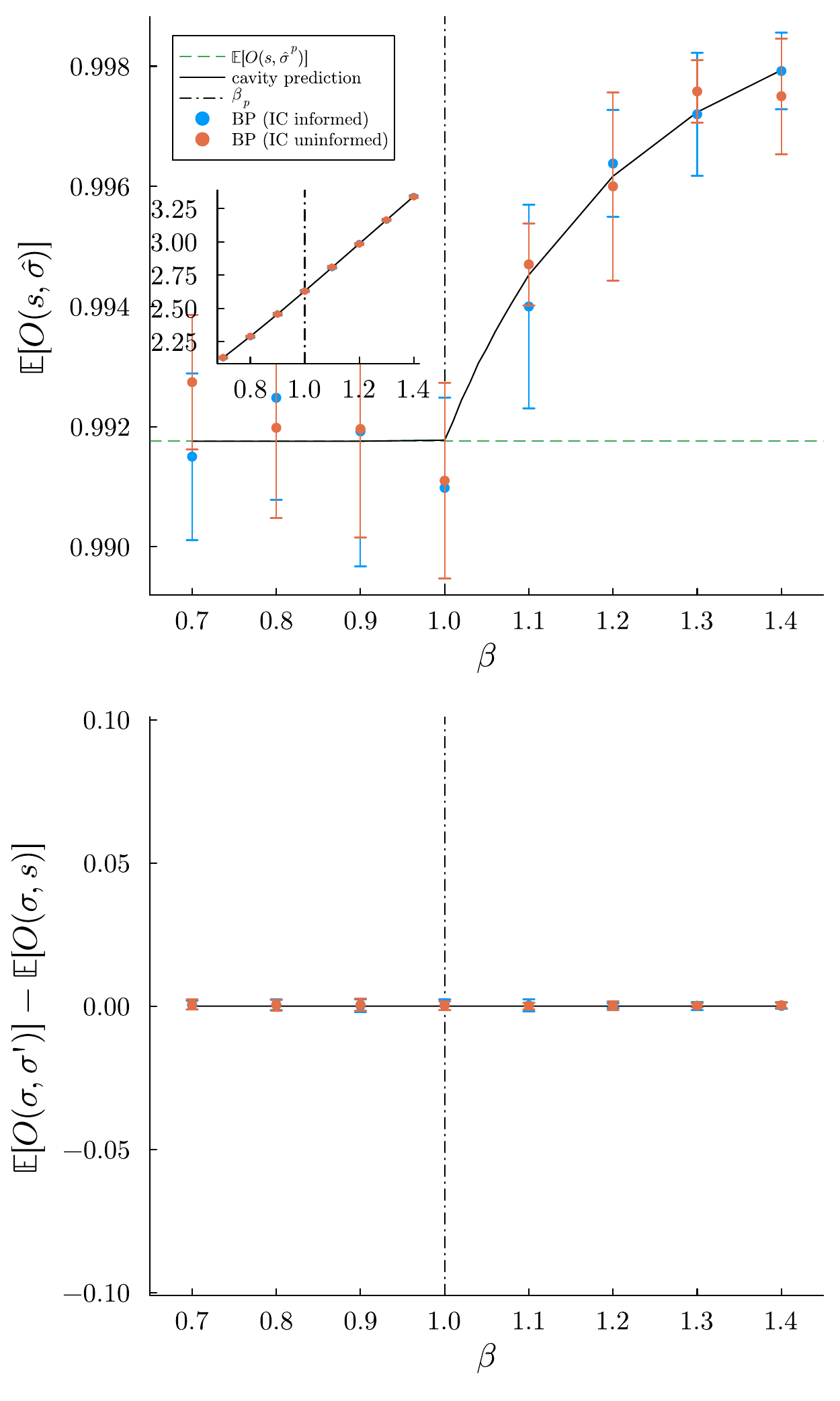}
			\caption{Results with ferromagnetic prior. Left: $\kappa=0.7$, Right: $\kappa=1.0$}
			\label{fig:supp_plot_ferro}
		\end{figure}
		\begin{figure}
			\centering
			\includegraphics[width=0.7\linewidth]{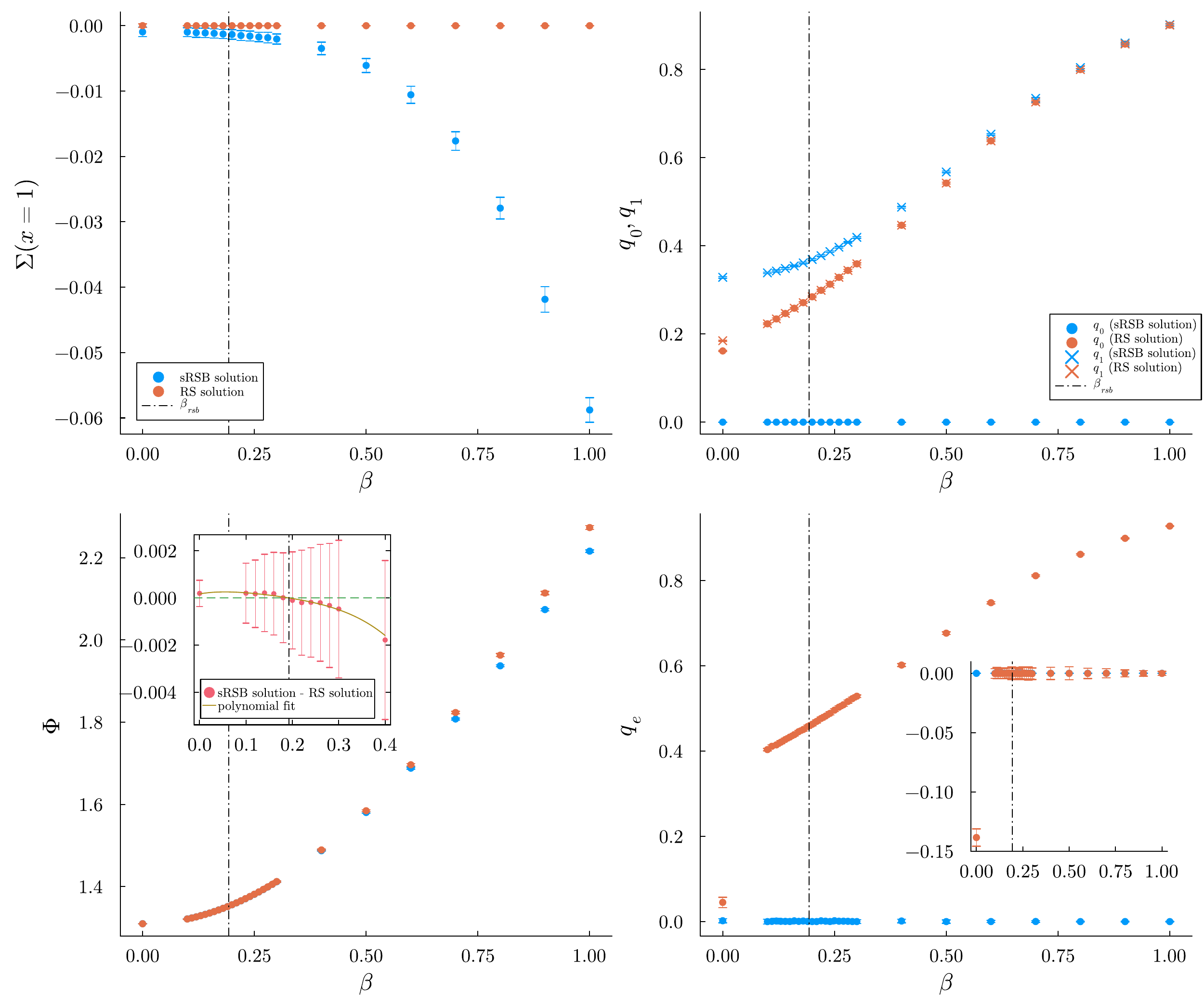}	
			\caption{Results with sRSB prior. $\kappa=-0.97$}
			\label{fig:supp_plot_sRSB_1}
		\end{figure}
		\begin{figure}
			\centering
			\includegraphics[width=0.7\linewidth]{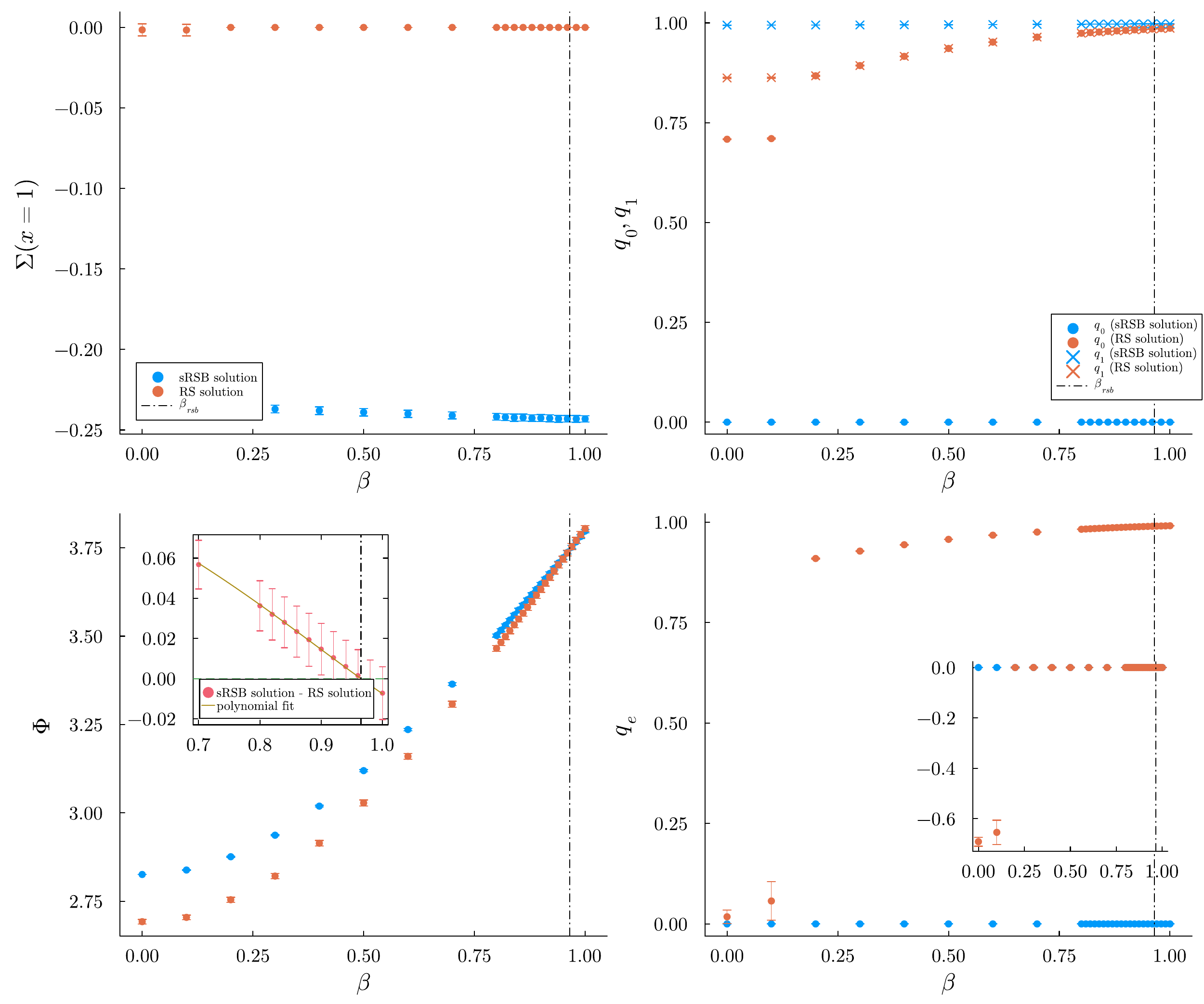}	
			\caption{Results with sRSB prior. $\kappa=-2.1$}
			\label{fig:supp_plot_sRSB_2}
		\end{figure}
	\end{center}
\end{widetext}

\end{document}